\documentclass[longauth]{aaEC}

\usepackage{graphicx}
\usepackage{natbib}
\usepackage{scalerel}
\usepackage{lscape}
\usepackage[table]{xcolor}

\usepackage{txfonts}
\usepackage[pdfencoding=auto,psdextra]{hyperref}
\hypersetup{
    colorlinks=true,
    linkcolor=blue,
    filecolor=magenta,      
    urlcolor=blue,
    citecolor=blue
}
\makeatletter
\renewcommand*\aa@pageof{, page \thepage{} of 11}
\makeatother

\usepackage[utf8]{inputenc}

\usepackage[switch, modulo]{lineno}
              
\nolinenumbers

\renewcommand{\linenumbers}[0]{}
\newcommand{\software}[1]{{\texttt{#1}}}
\usepackage{euclid}

\begin{document}
\title{\Euclid: Quick Data Release (Q1) -- Optical hotspots in powerful radio galaxies\thanks{This paper is published on behalf of the Euclid Consortium.}} 
\newcommand{\orcid}[1]{} 
\author{M.~J.~Hardcastle\orcid{0000-0003-4223-1117}\thanks{\email{m.j.hardcastle@herts.ac.uk}}\inst{\ref{aff1}}
\and L.~Bisigello\orcid{0000-0003-0492-4924}\inst{\ref{aff2}}
\and M.~Bondi\orcid{0000-0002-9553-7999}\inst{\ref{aff3}}
\and M.~Magliocchetti\orcid{0000-0001-9158-4838}\inst{\ref{aff4}}
\and L.~K.~Morabito\orcid{0000-0003-0487-6651}\inst{\ref{aff5},\ref{aff6}}
\and J.~Petley\orcid{0000-0002-4496-0754}\inst{\ref{aff7}}
\and F.~Ricci\orcid{0000-0001-5742-5980}\inst{\ref{aff8},\ref{aff9}}
\and H.~J.~A.~Rottgering\orcid{0000-0001-8887-2257}\inst{\ref{aff7}}
\and K.~Rubinur\orcid{0000-0001-5574-5104}\inst{\ref{aff10}}
\and D.~J.~B.~Smith\orcid{0000-0001-9708-253X}\inst{\ref{aff1}}
\and D.~Stern\orcid{0000-0003-2686-9241}\inst{\ref{aff11}}
\and B.~Altieri\orcid{0000-0003-3936-0284}\inst{\ref{aff12}}
\and S.~Andreon\orcid{0000-0002-2041-8784}\inst{\ref{aff13}}
\and N.~Auricchio\orcid{0000-0003-4444-8651}\inst{\ref{aff14}}
\and C.~Baccigalupi\orcid{0000-0002-8211-1630}\inst{\ref{aff15},\ref{aff16},\ref{aff17},\ref{aff18}}
\and M.~Baldi\orcid{0000-0003-4145-1943}\inst{\ref{aff19},\ref{aff14},\ref{aff20}}
\and A.~Balestra\orcid{0000-0002-6967-261X}\inst{\ref{aff2}}
\and S.~Bardelli\orcid{0000-0002-8900-0298}\inst{\ref{aff14}}
\and P.~Battaglia\orcid{0000-0002-7337-5909}\inst{\ref{aff14}}
\and A.~Biviano\orcid{0000-0002-0857-0732}\inst{\ref{aff16},\ref{aff15}}
\and E.~Branchini\orcid{0000-0002-0808-6908}\inst{\ref{aff21},\ref{aff22},\ref{aff13}}
\and M.~Brescia\orcid{0000-0001-9506-5680}\inst{\ref{aff23},\ref{aff24}}
\and S.~Camera\orcid{0000-0003-3399-3574}\inst{\ref{aff25},\ref{aff26},\ref{aff27}}
\and V.~Capobianco\orcid{0000-0002-3309-7692}\inst{\ref{aff27}}
\and C.~Carbone\orcid{0000-0003-0125-3563}\inst{\ref{aff28}}
\and J.~Carretero\orcid{0000-0002-3130-0204}\inst{\ref{aff29},\ref{aff30}}
\and S.~Casas\orcid{0000-0002-4751-5138}\inst{\ref{aff31},\ref{aff32}}
\and M.~Castellano\orcid{0000-0001-9875-8263}\inst{\ref{aff9}}
\and G.~Castignani\orcid{0000-0001-6831-0687}\inst{\ref{aff14}}
\and S.~Cavuoti\orcid{0000-0002-3787-4196}\inst{\ref{aff24},\ref{aff33}}
\and K.~C.~Chambers\orcid{0000-0001-6965-7789}\inst{\ref{aff34}}
\and A.~Cimatti\inst{\ref{aff35}}
\and C.~Colodro-Conde\inst{\ref{aff36}}
\and G.~Congedo\orcid{0000-0003-2508-0046}\inst{\ref{aff37}}
\and C.~J.~Conselice\orcid{0000-0003-1949-7638}\inst{\ref{aff38}}
\and L.~Conversi\orcid{0000-0002-6710-8476}\inst{\ref{aff39},\ref{aff12}}
\and Y.~Copin\orcid{0000-0002-5317-7518}\inst{\ref{aff40}}
\and A.~Costille\inst{\ref{aff41}}
\and F.~Courbin\orcid{0000-0003-0758-6510}\inst{\ref{aff42},\ref{aff43},\ref{aff44}}
\and H.~M.~Courtois\orcid{0000-0003-0509-1776}\inst{\ref{aff45}}
\and M.~Cropper\orcid{0000-0003-4571-9468}\inst{\ref{aff46}}
\and J.-G.~Cuby\orcid{0000-0002-8767-1442}\inst{\ref{aff47},\ref{aff41}}
\and A.~Da~Silva\orcid{0000-0002-6385-1609}\inst{\ref{aff48},\ref{aff49}}
\and H.~Degaudenzi\orcid{0000-0002-5887-6799}\inst{\ref{aff50}}
\and G.~De~Lucia\orcid{0000-0002-6220-9104}\inst{\ref{aff16}}
\and C.~Dolding\orcid{0009-0003-7199-6108}\inst{\ref{aff46}}
\and H.~Dole\orcid{0000-0002-9767-3839}\inst{\ref{aff51}}
\and F.~Dubath\orcid{0000-0002-6533-2810}\inst{\ref{aff50}}
\and X.~Dupac\inst{\ref{aff12}}
\and M.~Farina\orcid{0000-0002-3089-7846}\inst{\ref{aff4}}
\and R.~Farinelli\inst{\ref{aff14}}
\and S.~Ferriol\inst{\ref{aff40}}
\and P.~Fosalba\orcid{0000-0002-1510-5214}\inst{\ref{aff52},\ref{aff53}}
\and S.~Fotopoulou\orcid{0000-0002-9686-254X}\inst{\ref{aff54}}
\and M.~Frailis\orcid{0000-0002-7400-2135}\inst{\ref{aff16}}
\and E.~Franceschi\orcid{0000-0002-0585-6591}\inst{\ref{aff14}}
\and M.~Fumana\orcid{0000-0001-6787-5950}\inst{\ref{aff28}}
\and S.~Galeotta\orcid{0000-0002-3748-5115}\inst{\ref{aff16}}
\and K.~George\orcid{0000-0002-1734-8455}\inst{\ref{aff55}}
\and B.~Gillis\orcid{0000-0002-4478-1270}\inst{\ref{aff37}}
\and C.~Giocoli\orcid{0000-0002-9590-7961}\inst{\ref{aff14},\ref{aff20}}
\and J.~Gracia-Carpio\orcid{0000-0003-4689-3134}\inst{\ref{aff56}}
\and A.~Grazian\orcid{0000-0002-5688-0663}\inst{\ref{aff2}}
\and F.~Grupp\inst{\ref{aff56},\ref{aff57}}
\and S.~V.~H.~Haugan\orcid{0000-0001-9648-7260}\inst{\ref{aff10}}
\and J.~Hoar\inst{\ref{aff12}}
\and W.~Holmes\inst{\ref{aff11}}
\and I.~M.~Hook\orcid{0000-0002-2960-978X}\inst{\ref{aff58}}
\and F.~Hormuth\inst{\ref{aff59}}
\and A.~Hornstrup\orcid{0000-0002-3363-0936}\inst{\ref{aff60},\ref{aff61}}
\and K.~Jahnke\orcid{0000-0003-3804-2137}\inst{\ref{aff62}}
\and M.~Jhabvala\inst{\ref{aff63}}
\and B.~Joachimi\orcid{0000-0001-7494-1303}\inst{\ref{aff64}}
\and S.~Kermiche\orcid{0000-0002-0302-5735}\inst{\ref{aff65}}
\and A.~Kiessling\orcid{0000-0002-2590-1273}\inst{\ref{aff11}}
\and B.~Kubik\orcid{0009-0006-5823-4880}\inst{\ref{aff40}}
\and M.~K\"ummel\orcid{0000-0003-2791-2117}\inst{\ref{aff57}}
\and M.~Kunz\orcid{0000-0002-3052-7394}\inst{\ref{aff66}}
\and H.~Kurki-Suonio\orcid{0000-0002-4618-3063}\inst{\ref{aff67},\ref{aff68}}
\and A.~M.~C.~Le~Brun\orcid{0000-0002-0936-4594}\inst{\ref{aff69}}
\and S.~Ligori\orcid{0000-0003-4172-4606}\inst{\ref{aff27}}
\and P.~B.~Lilje\orcid{0000-0003-4324-7794}\inst{\ref{aff10}}
\and V.~Lindholm\orcid{0000-0003-2317-5471}\inst{\ref{aff67},\ref{aff68}}
\and I.~Lloro\orcid{0000-0001-5966-1434}\inst{\ref{aff70}}
\and G.~Mainetti\orcid{0000-0003-2384-2377}\inst{\ref{aff71}}
\and O.~Mansutti\orcid{0000-0001-5758-4658}\inst{\ref{aff16}}
\and O.~Marggraf\orcid{0000-0001-7242-3852}\inst{\ref{aff72}}
\and M.~Martinelli\orcid{0000-0002-6943-7732}\inst{\ref{aff9},\ref{aff73}}
\and N.~Martinet\orcid{0000-0003-2786-7790}\inst{\ref{aff41}}
\and F.~Marulli\orcid{0000-0002-8850-0303}\inst{\ref{aff74},\ref{aff14},\ref{aff20}}
\and R.~J.~Massey\orcid{0000-0002-6085-3780}\inst{\ref{aff6}}
\and E.~Medinaceli\orcid{0000-0002-4040-7783}\inst{\ref{aff14}}
\and S.~Mei\orcid{0000-0002-2849-559X}\inst{\ref{aff75},\ref{aff76}}
\and M.~Meneghetti\orcid{0000-0003-1225-7084}\inst{\ref{aff14},\ref{aff20}}
\and E.~Merlin\orcid{0000-0001-6870-8900}\inst{\ref{aff9}}
\and G.~Meylan\inst{\ref{aff77}}
\and A.~Mora\orcid{0000-0002-1922-8529}\inst{\ref{aff78}}
\and M.~Moresco\orcid{0000-0002-7616-7136}\inst{\ref{aff74},\ref{aff14}}
\and L.~Moscardini\orcid{0000-0002-3473-6716}\inst{\ref{aff74},\ref{aff14},\ref{aff20}}
\and R.~Nakajima\orcid{0009-0009-1213-7040}\inst{\ref{aff72}}
\and C.~Neissner\orcid{0000-0001-8524-4968}\inst{\ref{aff79},\ref{aff30}}
\and S.-M.~Niemi\orcid{0009-0005-0247-0086}\inst{\ref{aff80}}
\and J.~W.~Nightingale\orcid{0000-0002-8987-7401}\inst{\ref{aff81}}
\and C.~Padilla\orcid{0000-0001-7951-0166}\inst{\ref{aff79}}
\and S.~Paltani\orcid{0000-0002-8108-9179}\inst{\ref{aff50}}
\and F.~Pasian\orcid{0000-0002-4869-3227}\inst{\ref{aff16}}
\and K.~Pedersen\inst{\ref{aff82}}
\and W.~J.~Percival\orcid{0000-0002-0644-5727}\inst{\ref{aff83},\ref{aff84},\ref{aff85}}
\and V.~Pettorino\orcid{0000-0002-4203-9320}\inst{\ref{aff80}}
\and S.~Pires\orcid{0000-0002-0249-2104}\inst{\ref{aff86}}
\and G.~Polenta\orcid{0000-0003-4067-9196}\inst{\ref{aff87}}
\and L.~A.~Popa\inst{\ref{aff88}}
\and F.~Raison\orcid{0000-0002-7819-6918}\inst{\ref{aff56}}
\and A.~Renzi\orcid{0000-0001-9856-1970}\inst{\ref{aff89},\ref{aff90},\ref{aff14}}
\and J.~Rhodes\orcid{0000-0002-4485-8549}\inst{\ref{aff11}}
\and G.~Riccio\inst{\ref{aff24}}
\and E.~Romelli\orcid{0000-0003-3069-9222}\inst{\ref{aff16}}
\and M.~Roncarelli\orcid{0000-0001-9587-7822}\inst{\ref{aff14}}
\and B.~Rusholme\orcid{0000-0001-7648-4142}\inst{\ref{aff91}}
\and R.~Saglia\orcid{0000-0003-0378-7032}\inst{\ref{aff57},\ref{aff56}}
\and Z.~Sakr\orcid{0000-0002-4823-3757}\inst{\ref{aff92},\ref{aff93},\ref{aff94}}
\and D.~Sapone\orcid{0000-0001-7089-4503}\inst{\ref{aff95}}
\and B.~Sartoris\orcid{0000-0003-1337-5269}\inst{\ref{aff57},\ref{aff16}}
\and M.~Schirmer\orcid{0000-0003-2568-9994}\inst{\ref{aff62}}
\and P.~Schneider\orcid{0000-0001-8561-2679}\inst{\ref{aff72}}
\and A.~Secroun\orcid{0000-0003-0505-3710}\inst{\ref{aff65}}
\and G.~Seidel\orcid{0000-0003-2907-353X}\inst{\ref{aff62}}
\and E.~Sihvola\orcid{0000-0003-1804-7715}\inst{\ref{aff96}}
\and P.~Simon\inst{\ref{aff72}}
\and C.~Sirignano\orcid{0000-0002-0995-7146}\inst{\ref{aff89},\ref{aff90}}
\and G.~Sirri\orcid{0000-0003-2626-2853}\inst{\ref{aff20}}
\and J.~Skottfelt\orcid{0000-0003-1310-8283}\inst{\ref{aff97}}
\and L.~Stanco\orcid{0000-0002-9706-5104}\inst{\ref{aff90}}
\and P.~Tallada-Cresp\'{i}\orcid{0000-0002-1336-8328}\inst{\ref{aff29},\ref{aff30}}
\and A.~N.~Taylor\inst{\ref{aff37}}
\and H.~I.~Teplitz\orcid{0000-0002-7064-5424}\inst{\ref{aff98}}
\and I.~Tereno\orcid{0000-0002-4537-6218}\inst{\ref{aff48},\ref{aff99}}
\and N.~Tessore\orcid{0000-0002-9696-7931}\inst{\ref{aff46}}
\and S.~Toft\orcid{0000-0003-3631-7176}\inst{\ref{aff100},\ref{aff101}}
\and R.~Toledo-Moreo\orcid{0000-0002-2997-4859}\inst{\ref{aff102}}
\and F.~Torradeflot\orcid{0000-0003-1160-1517}\inst{\ref{aff30},\ref{aff29}}
\and I.~Tutusaus\orcid{0000-0002-3199-0399}\inst{\ref{aff53},\ref{aff52},\ref{aff93}}
\and J.~Valiviita\orcid{0000-0001-6225-3693}\inst{\ref{aff67},\ref{aff68}}
\and T.~Vassallo\orcid{0000-0001-6512-6358}\inst{\ref{aff16},\ref{aff55}}
\and G.~Verdoes~Kleijn\orcid{0000-0001-5803-2580}\inst{\ref{aff103}}
\and A.~Veropalumbo\orcid{0000-0003-2387-1194}\inst{\ref{aff13},\ref{aff22},\ref{aff21}}
\and Y.~Wang\orcid{0000-0002-4749-2984}\inst{\ref{aff91}}
\and J.~Weller\orcid{0000-0002-8282-2010}\inst{\ref{aff57},\ref{aff56}}
\and O.~R.~Williams\orcid{0000-0003-0274-1526}\inst{\ref{aff104}}
\and A.~Zacchei\orcid{0000-0003-0396-1192}\inst{\ref{aff16},\ref{aff15}}
\and G.~Zamorani\orcid{0000-0002-2318-301X}\inst{\ref{aff14}}
\and F.~M.~Zerbi\orcid{0000-0002-9996-973X}\inst{\ref{aff13}}
\and E.~Zucca\orcid{0000-0002-5845-8132}\inst{\ref{aff14}}
\and J.~Garc\'ia-Bellido\orcid{0000-0002-9370-8360}\inst{\ref{aff92}}
\and V.~Scottez\orcid{0009-0008-3864-940X}\inst{\ref{aff105},\ref{aff106}}
\and F.~Tarsitano\orcid{0000-0002-5919-0238}\inst{\ref{aff107},\ref{aff108},\ref{aff50}}
\and J.~H.~Knapen\orcid{0000-0003-1643-0024}\inst{\ref{aff36},\ref{aff109}}}
										   
\institute{Department of Physics, Astronomy and Mathematics, University of Hertfordshire, College Lane, Hatfield AL10 9AB, UK\label{aff1}
\and
INAF-Osservatorio Astronomico di Padova, Via dell'Osservatorio 5, 35122 Padova, Italy\label{aff2}
\and
INAF, Istituto di Radioastronomia, Via Piero Gobetti 101, 40129 Bologna, Italy\label{aff3}
\and
INAF-Istituto di Astrofisica e Planetologia Spaziali, via del Fosso del Cavaliere, 100, 00100 Roma, Italy\label{aff4}
\and
Department of Physics, Centre for Extragalactic Astronomy, Durham University, South Road, Durham, DH1 3LE, UK\label{aff5}
\and
Department of Physics, Institute for Computational Cosmology, Durham University, South Road, Durham, DH1 3LE, UK\label{aff6}
\and
Leiden Observatory, Leiden University, Einsteinweg 55, 2333 CC Leiden, The Netherlands\label{aff7}
\and
Department of Mathematics and Physics, Roma Tre University, Via della Vasca Navale 84, 00146 Rome, Italy\label{aff8}
\and
INAF-Osservatorio Astronomico di Roma, Via Frascati 33, 00078 Monteporzio Catone, Italy\label{aff9}
\and
Institute of Theoretical Astrophysics, University of Oslo, P.O. Box 1029 Blindern, 0315 Oslo, Norway\label{aff10}
\and
Jet Propulsion Laboratory, California Institute of Technology, 4800 Oak Grove Drive, Pasadena, CA, 91109, USA\label{aff11}
\and
ESAC/ESA, Camino Bajo del Castillo, s/n., Urb. Villafranca del Castillo, 28692 Villanueva de la Ca\~nada, Madrid, Spain\label{aff12}
\and
INAF-Osservatorio Astronomico di Brera, Via Brera 28, 20122 Milano, Italy\label{aff13}
\and
INAF-Osservatorio di Astrofisica e Scienza dello Spazio di Bologna, Via Piero Gobetti 93/3, 40129 Bologna, Italy\label{aff14}
\and
IFPU, Institute for Fundamental Physics of the Universe, via Beirut 2, 34151 Trieste, Italy\label{aff15}
\and
INAF-Osservatorio Astronomico di Trieste, Via G. B. Tiepolo 11, 34143 Trieste, Italy\label{aff16}
\and
INFN, Sezione di Trieste, Via Valerio 2, 34127 Trieste TS, Italy\label{aff17}
\and
SISSA, International School for Advanced Studies, Via Bonomea 265, 34136 Trieste TS, Italy\label{aff18}
\and
Dipartimento di Fisica e Astronomia, Universit\`a di Bologna, Via Gobetti 93/2, 40129 Bologna, Italy\label{aff19}
\and
INFN-Sezione di Bologna, Viale Berti Pichat 6/2, 40127 Bologna, Italy\label{aff20}
\and
Dipartimento di Fisica, Universit\`a di Genova, Via Dodecaneso 33, 16146, Genova, Italy\label{aff21}
\and
INFN-Sezione di Genova, Via Dodecaneso 33, 16146, Genova, Italy\label{aff22}
\and
Department of Physics "E. Pancini", University Federico II, Via Cinthia 6, 80126, Napoli, Italy\label{aff23}
\and
INAF-Osservatorio Astronomico di Capodimonte, Via Moiariello 16, 80131 Napoli, Italy\label{aff24}
\and
Dipartimento di Fisica, Universit\`a degli Studi di Torino, Via P. Giuria 1, 10125 Torino, Italy\label{aff25}
\and
INFN-Sezione di Torino, Via P. Giuria 1, 10125 Torino, Italy\label{aff26}
\and
INAF-Osservatorio Astrofisico di Torino, Via Osservatorio 20, 10025 Pino Torinese (TO), Italy\label{aff27}
\and
INAF-IASF Milano, Via Alfonso Corti 12, 20133 Milano, Italy\label{aff28}
\and
Centro de Investigaciones Energ\'eticas, Medioambientales y Tecnol\'ogicas (CIEMAT), Avenida Complutense 40, 28040 Madrid, Spain\label{aff29}
\and
Port d'Informaci\'{o} Cient\'{i}fica, Campus UAB, C. Albareda s/n, 08193 Bellaterra (Barcelona), Spain\label{aff30}
\and
Institute for Theoretical Particle Physics and Cosmology (TTK), RWTH Aachen University, 52056 Aachen, Germany\label{aff31}
\and
Deutsches Zentrum f\"ur Luft- und Raumfahrt e. V. (DLR), Linder H\"ohe, 51147 K\"oln, Germany\label{aff32}
\and
INFN section of Naples, Via Cinthia 6, 80126, Napoli, Italy\label{aff33}
\and
Institute for Astronomy, University of Hawaii, 2680 Woodlawn Drive, Honolulu, HI 96822, USA\label{aff34}
\and
Dipartimento di Fisica e Astronomia "Augusto Righi" - Alma Mater Studiorum Universit\`a di Bologna, Viale Berti Pichat 6/2, 40127 Bologna, Italy\label{aff35}
\and
Instituto de Astrof\'{\i}sica de Canarias, E-38205 La Laguna, Tenerife, Spain\label{aff36}
\and
Institute for Astronomy, University of Edinburgh, Royal Observatory, Blackford Hill, Edinburgh EH9 3HJ, UK\label{aff37}
\and
Jodrell Bank Centre for Astrophysics, Department of Physics and Astronomy, University of Manchester, Oxford Road, Manchester M13 9PL, UK\label{aff38}
\and
European Space Agency/ESRIN, Largo Galileo Galilei 1, 00044 Frascati, Roma, Italy\label{aff39}
\and
Universit\'e Claude Bernard Lyon 1, CNRS/IN2P3, IP2I Lyon, UMR 5822, Villeurbanne, F-69100, France\label{aff40}
\and
Aix-Marseille Universit\'e, CNRS, CNES, LAM, Marseille, France\label{aff41}
\and
Institut de Ci\`{e}ncies del Cosmos (ICCUB), Universitat de Barcelona (IEEC-UB), Mart\'{i} i Franqu\`{e}s 1, 08028 Barcelona, Spain\label{aff42}
\and
Instituci\'o Catalana de Recerca i Estudis Avan\c{c}ats (ICREA), Passeig de Llu\'{\i}s Companys 23, 08010 Barcelona, Spain\label{aff43}
\and
Institut de Ciencies de l'Espai (IEEC-CSIC), Campus UAB, Carrer de Can Magrans, s/n Cerdanyola del Vall\'es, 08193 Barcelona, Spain\label{aff44}
\and
UCB Lyon 1, CNRS/IN2P3, IUF, IP2I Lyon, 4 rue Enrico Fermi, 69622 Villeurbanne, France\label{aff45}
\and
Mullard Space Science Laboratory, University College London, Holmbury St Mary, Dorking, Surrey RH5 6NT, UK\label{aff46}
\and
Canada-France-Hawaii Telescope, 65-1238 Mamalahoa Hwy, Kamuela, HI 96743, USA\label{aff47}
\and
Departamento de F\'isica, Faculdade de Ci\^encias, Universidade de Lisboa, Edif\'icio C8, Campo Grande, PT1749-016 Lisboa, Portugal\label{aff48}
\and
Instituto de Astrof\'isica e Ci\^encias do Espa\c{c}o, Faculdade de Ci\^encias, Universidade de Lisboa, Campo Grande, 1749-016 Lisboa, Portugal\label{aff49}
\and
Department of Astronomy, University of Geneva, ch. d'Ecogia 16, 1290 Versoix, Switzerland\label{aff50}
\and
Universit\'e Paris-Saclay, CNRS, Institut d'astrophysique spatiale, 91405, Orsay, France\label{aff51}
\and
Institut d'Estudis Espacials de Catalunya (IEEC),  Edifici RDIT, Campus UPC, 08860 Castelldefels, Barcelona, Spain\label{aff52}
\and
Institute of Space Sciences (ICE, CSIC), Campus UAB, Carrer de Can Magrans, s/n, 08193 Barcelona, Spain\label{aff53}
\and
School of Physics, HH Wills Physics Laboratory, University of Bristol, Tyndall Avenue, Bristol, BS8 1TL, UK\label{aff54}
\and
University Observatory, LMU Faculty of Physics, Scheinerstr.~1, 81679 Munich, Germany\label{aff55}
\and
Max Planck Institute for Extraterrestrial Physics, Giessenbachstr. 1, 85748 Garching, Germany\label{aff56}
\and
Universit\"ats-Sternwarte M\"unchen, Fakult\"at f\"ur Physik, Ludwig-Maximilians-Universit\"at M\"unchen, Scheinerstr.~1, 81679 M\"unchen, Germany\label{aff57}
\and
Department of Physics, Lancaster University, Lancaster, LA1 4YB, UK\label{aff58}
\and
Felix Hormuth Engineering, Goethestr. 17, 69181 Leimen, Germany\label{aff59}
\and
Technical University of Denmark, Elektrovej 327, 2800 Kgs. Lyngby, Denmark\label{aff60}
\and
Cosmic Dawn Center (DAWN), Denmark\label{aff61}
\and
Max-Planck-Institut f\"ur Astronomie, K\"onigstuhl 17, 69117 Heidelberg, Germany\label{aff62}
\and
NASA Goddard Space Flight Center, Greenbelt, MD 20771, USA\label{aff63}
\and
Department of Physics and Astronomy, University College London, Gower Street, London WC1E 6BT, UK\label{aff64}
\and
Aix-Marseille Universit\'e, CNRS/IN2P3, CPPM, Marseille, France\label{aff65}
\and
Universit\'e de Gen\`eve, D\'epartement de Physique Th\'eorique and Centre for Astroparticle Physics, 24 quai Ernest-Ansermet, CH-1211 Gen\`eve 4, Switzerland\label{aff66}
\and
Department of Physics, P.O. Box 64, University of Helsinki, 00014 Helsinki, Finland\label{aff67}
\and
Helsinki Institute of Physics, Gustaf H{\"a}llstr{\"o}min katu 2, University of Helsinki, 00014 Helsinki, Finland\label{aff68}
\and
Laboratoire d'etude de l'Univers et des phenomenes eXtremes, Observatoire de Paris, Universit\'e PSL, Sorbonne Universit\'e, CNRS, 92190 Meudon, France\label{aff69}
\and
SKAO, Jodrell Bank, Lower Withington, Macclesfield SK11 9FT, UK\label{aff70}
\and
Centre de Calcul de l'IN2P3/CNRS, 21 avenue Pierre de Coubertin 69627 Villeurbanne Cedex, France\label{aff71}
\and
Universit\"at Bonn, Argelander-Institut f\"ur Astronomie, Auf dem H\"ugel 71, 53121 Bonn, Germany\label{aff72}
\and
INFN-Sezione di Roma, Piazzale Aldo Moro, 2 - c/o Dipartimento di Fisica, Edificio G. Marconi, 00185 Roma, Italy\label{aff73}
\and
Dipartimento di Fisica e Astronomia "Augusto Righi" - Alma Mater Studiorum Universit\`a di Bologna, via Piero Gobetti 93/2, 40129 Bologna, Italy\label{aff74}
\and
Universit\'e Paris Cit\'e, CNRS, Astroparticule et Cosmologie, 75013 Paris, France\label{aff75}
\and
CNRS-UCB International Research Laboratory, Centre Pierre Bin\'etruy, IRL2007, CPB-IN2P3, Berkeley, USA\label{aff76}
\and
Institute of Physics, Laboratory of Astrophysics, Ecole Polytechnique F\'ed\'erale de Lausanne (EPFL), Observatoire de Sauverny, 1290 Versoix, Switzerland\label{aff77}
\and
Telespazio UK S.L. for European Space Agency (ESA), Camino bajo del Castillo, s/n, Urbanizacion Villafranca del Castillo, Villanueva de la Ca\~nada, 28692 Madrid, Spain\label{aff78}
\and
Institut de F\'{i}sica d'Altes Energies (IFAE), The Barcelona Institute of Science and Technology, Campus UAB, 08193 Bellaterra (Barcelona), Spain\label{aff79}
\and
European Space Agency/ESTEC, Keplerlaan 1, 2201 AZ Noordwijk, The Netherlands\label{aff80}
\and
School of Mathematics, Statistics and Physics, Newcastle University, Herschel Building, Newcastle-upon-Tyne, NE1 7RU, UK\label{aff81}
\and
DARK, Niels Bohr Institute, University of Copenhagen, Jagtvej 155, 2200 Copenhagen, Denmark\label{aff82}
\and
Waterloo Centre for Astrophysics, University of Waterloo, Waterloo, Ontario N2L 3G1, Canada\label{aff83}
\and
Department of Physics and Astronomy, University of Waterloo, Waterloo, Ontario N2L 3G1, Canada\label{aff84}
\and
Perimeter Institute for Theoretical Physics, Waterloo, Ontario N2L 2Y5, Canada\label{aff85}
\and
Universit\'e Paris-Saclay, Universit\'e Paris Cit\'e, CEA, CNRS, AIM, 91191, Gif-sur-Yvette, France\label{aff86}
\and
Space Science Data Center, Italian Space Agency, via del Politecnico snc, 00133 Roma, Italy\label{aff87}
\and
Institute of Space Science, Str. Atomistilor, nr. 409 M\u{a}gurele, Ilfov, 077125, Romania\label{aff88}
\and
Dipartimento di Fisica e Astronomia "G. Galilei", Universit\`a di Padova, Via Marzolo 8, 35131 Padova, Italy\label{aff89}
\and
INFN-Padova, Via Marzolo 8, 35131 Padova, Italy\label{aff90}
\and
Caltech/IPAC, 1200 E. California Blvd., Pasadena, CA 91125, USA\label{aff91}
\and
Instituto de F\'isica Te\'orica UAM-CSIC, Campus de Cantoblanco, 28049 Madrid, Spain\label{aff92}
\and
Institut de Recherche en Astrophysique et Plan\'etologie (IRAP), Universit\'e de Toulouse, CNRS, UPS, CNES, 14 Av. Edouard Belin, 31400 Toulouse, France\label{aff93}
\and
Universit\'e St Joseph; Faculty of Sciences, Beirut, Lebanon\label{aff94}
\and
Departamento de F\'isica, FCFM, Universidad de Chile, Blanco Encalada 2008, Santiago, Chile\label{aff95}
\and
Department of Physics and Helsinki Institute of Physics, Gustaf H\"allstr\"omin katu 2, University of Helsinki, 00014 Helsinki, Finland\label{aff96}
\and
Centre for Electronic Imaging, Open University, Walton Hall, Milton Keynes, MK7~6AA, UK\label{aff97}
\and
Infrared Processing and Analysis Center, California Institute of Technology, Pasadena, CA 91125, USA\label{aff98}
\and
Instituto de Astrof\'isica e Ci\^encias do Espa\c{c}o, Faculdade de Ci\^encias, Universidade de Lisboa, Tapada da Ajuda, 1349-018 Lisboa, Portugal\label{aff99}
\and
Cosmic Dawn Center (DAWN)\label{aff100}
\and
Niels Bohr Institute, University of Copenhagen, Jagtvej 128, 2200 Copenhagen, Denmark\label{aff101}
\and
Universidad Polit\'ecnica de Cartagena, Departamento de Electr\'onica y Tecnolog\'ia de Computadoras,  Plaza del Hospital 1, 30202 Cartagena, Spain\label{aff102}
\and
Kapteyn Astronomical Institute, University of Groningen, PO Box 800, 9700 AV Groningen, The Netherlands\label{aff103}
\and
Centre for Information Technology, University of Groningen, P.O. Box 11044, 9700 CA Groningen, The Netherlands\label{aff104}
\and
Institut d'Astrophysique de Paris, 98bis Boulevard Arago, 75014, Paris, France\label{aff105}
\and
ICL, Junia, Universit\'e Catholique de Lille, LITL, 59000 Lille, France\label{aff106}
\and
Kobayashi-Maskawa Institute for the Origin of Particles and the Universe, Nagoya University, Chikusa-ku, Nagoya, 464-8602, Japan\label{aff107}
\and
Institute for Particle Physics and Astrophysics, Dept. of Physics, ETH Zurich, Wolfgang-Pauli-Strasse 27, 8093 Zurich, Switzerland\label{aff108}
\and
Universidad de La Laguna, Dpto. Astrof\'\i sica, E-38206 La Laguna, Tenerife, Spain\label{aff109}}
%
%
\abstract{ The hotspots of powerful radio galaxies are the sites of
  high-energy particle acceleration, but radio observations give us
  limited information on the electron energy distributions that they
  generate, which affects our overall understanding of their physics
  and energetics. Optical synchrotron counterparts to radio galaxy
  hotspots, which probe high electron energies, have been known for
  many years, but have only been studied in a small number of objects
  selected as being very bright in the radio. By combining
  Low-Frequency Array (LOFAR) observations with the Euclid Q1 data
  release in the Euclid Deep Field North, we identify 77 powerful
  radio galaxies with hotspots and show that a small but significant
  fraction of them, 7--15\%, may have detected infrared and optical
  counterparts to their hotspots. The hotspot spectral energy
  distributions in our clearest cases are consistent with particle
  acceleration up to (but not beyond) 1\,TeV in energy on the
  assumption of field strengths close to equipartition, whereas
  non-detected hotspots either do not reach such high electron
  energies or would have to have very different electron energy
  spectra or significantly sub-equipartition magnetic field strengths.
  Two new candidate optical hotspots are clearly detected matching
  high-resolution LOFAR data; one is remarkable as both the
  highest-redshift optical hotspot known to date ($z \approx 1.6$) and
  also the first optical detection of the hotspot of the inner lobe of
  a restarting source. We discuss approaches to performing a
  large-scale study of hotspots with the full wide-area
  \Euclid\ surveys, particularly when combined with the international
  baselines of LOFAR, enabling for the first time a statistical
  analysis of high-energy particle acceleration in hotspots. }
   \titlerunning{Euclid Q1: Optical hotspots}
   \authorrunning{M.J. Hardcastle et al.}
   
   \maketitle
   
\section{Introduction}
\label{sec:intro}

The hotspots of Fanaroff--Riley class II (FRII) radio galaxies
\citep{Fanaroff+Riley74} are generally interpreted as being related to
the terminal shocks of the supersonic jets in this most powerful class
of radio-luminous active galactic nucleus (RLAGN). Their radio spectra
are consistent with being sites of particle acceleration
\citep[e.g.][]{Heavens+Meisenheimer87,Meisenheimer+89}. In general
the radio spectra of FRII lobes steepen with distance away from the
hotspots \citep[e.g.][]{Alexander+Leahy87,Harwood+17}, consistent with
the idea that injection of freshly accelerated particles takes place
at the hotspots. A complication of this picture is that
high-resolution radio observations of FRIIs show that hotspots rarely
have the appearance of simple planar shocks
\citep[e.g.][]{Lonsdale+Barthel86,Bridle+94,Leahy+97,Hardcastle+98}.
Multiple hotspots, or broad regions of high surface brightness with
many peaks, are common when these objects are observed in
detail, and it is often not clear which, if any, of the radio peaks
are the current site of particle acceleration. This ambiguity about
the location of the terminal shock in the radio makes it hard to test
any of the numerous models of multiple hotspot formation
\citep[e.g.][]{Scheuer82,Williams+Gull85,Horton+23}.

Key additional information is provided by observations of hotspots at
other frequencies. The synchrotron emission process is inherently
broad-band, and the existence of optical counterparts to hotspots
provided one of the earliest pieces of evidence for the model of
in situ particle acceleration at these locations
\citep{Saslaw+78}, since high-frequency synchrotron emission
corresponds to relativistic electrons with short radiative lifetimes
(hundreds to thousands of years for optical synchrotron) implying that
the radiating particles cannot be transported from the nucleus, which
is often hundreds of kpc away. Early direct evidence that the optical
emission was synchrotron in some well-studied cases, rather than
inverse-Compton emission or a chance superposition with a background
object, came from polarimetric observations
\citep[e.g][]{Meisenheimer+Roeser86,Lahteenmaki+Valtaoja99} showing
that the optical sources, like their radio counterparts, were highly
polarized. In some cases, both components of a double hotspot were
detected in optical synchrotron emission, challenging models in which
there is a single site of active particle acceleration in a lobe. More
recently, X-ray emission thought to be of synchrotron origin has been
detected in a number of powerful FRII sources, mostly using
\Chandra\ \citep[e.g.][]{Hardcastle+04,Jiminez-Gallardo+21} although
as yet no X-ray polarization has been detected from even the brightest
known hotspot source, Pic A \citep{Tugliani+25}, despite its likely
synchrotron origin \citep{Hardcastle+16}. Again, there are several
known cases where what appears to be X-ray synchrotron emission
originates from multiple discrete hotspots \citep{Hardcastle+07}, from
a broader diffuse region around them, or from a combination of both,
challenging the notion that particle acceleration takes place simply
at a single jet termination shock.

What is missing so far from multi-wavelength studies of hotspots is
statistical information about the population. Therefore, while, for
example, 
\cite{Meisenheimer+97} in the optical and infrared, or
\cite{Hardcastle+04} in the X-ray, noted that some hotspots in their
samples were detected while others were not, their work was based on
samples of a few to a few tens of sources, mostly bright objects from
the 3C catalogue, due to the requirement for expensive pointed optical
or X-ray observations. In addition, ground-based optical observations
suffer from relatively low resolution, which means that it is hard to
distinguish optical hotspots from background galaxies in the absence
of polarimetric measurements, which are hard to obtain.
\textit{Hubble} Space Telescope observations of samples of hundreds of
FRII sources \citep[e.g.][]{deKoff+96} generally have not been able to
provide constraints on optical hotspot properties due to HST's
small field of view. Consequently, we do not have a clear picture of
whether particle acceleration to high energies is a property
of hotspots in general, is associated with some particular feature of
a hotspot such as the current jet termination, or is simply a
transient consequence of the complex magnetohydrodynamical processes
taking place in the hotspot region. Only a large-scale, unbiased,
systematic study of the multi-wavelength properties of hotspots can
allow us to answer these questions.

The \Euclid\ mission \citep{EuclidSkyOverview} is carrying out a
wide-area survey of the extragalactic sky at infrared (IR) and optical
wavelengths with a resolution (FWHM in the range
\mbox{\ang{;;0.35}}--\mbox{\ang{;;0.38}} in the near-IR,
\mbox{\ang{;;0.16}} in the optical:
\citealt{Libralato24,EuclidSkyNISP,EuclidSkyVIS}) that is very well
matched to the spatial scale of hotspots in resolved RLAGN
\citep{Hardcastle+98}, and a $5\,\sigma$ point source sensitivity
around 0.7\,$\mu$Jy at IR wavelengths, which in principle could allow
it to detect optical and IR counterparts to radio hotspots that have
mJy flux densities at frequencies of 1\,GHz. \Euclid, combined with
sensitive and high-resolution radio observations, thus offers us a
chance to build up a systematic picture of high-energy particle
acceleration in these objects for the first time. In this paper we
combine the Euclid Q1 data \citep{Q1cite} from the Euclid Deep Field
North (EDF-N) with Low-Frequency Array (LOFAR) observations at
resolutions of 6\arcsecond\ and \mbox{\ang{;;0.5}}, to place the first
systematic constraints on high-energy particle acceleration in the
numerous population of FRII radio galaxies detected using LOFAR
surveys, which are typically of much lower luminosity or at higher
redshift than objects previously studied in this way.

Throughout the paper, we use a concordance cosmology in which $H_0 =
70$\,km\,s$^{-1}$\,Mpc$^{-1}$, $\Omega_\mathrm{m} = 0.3$, and $\Omega_\Lambda =
0.7$. Spectral index $\alpha$ is defined in the sense $S_\nu \propto
\nu^{-\alpha}$ and, for calculation of luminosities, we take the
typical spectral index of an extended radio source to be $\alpha =
0.7$ \citep{Conway+63}.

\section{Data}

\begin{figure}
  \includegraphics[width=\linewidth]{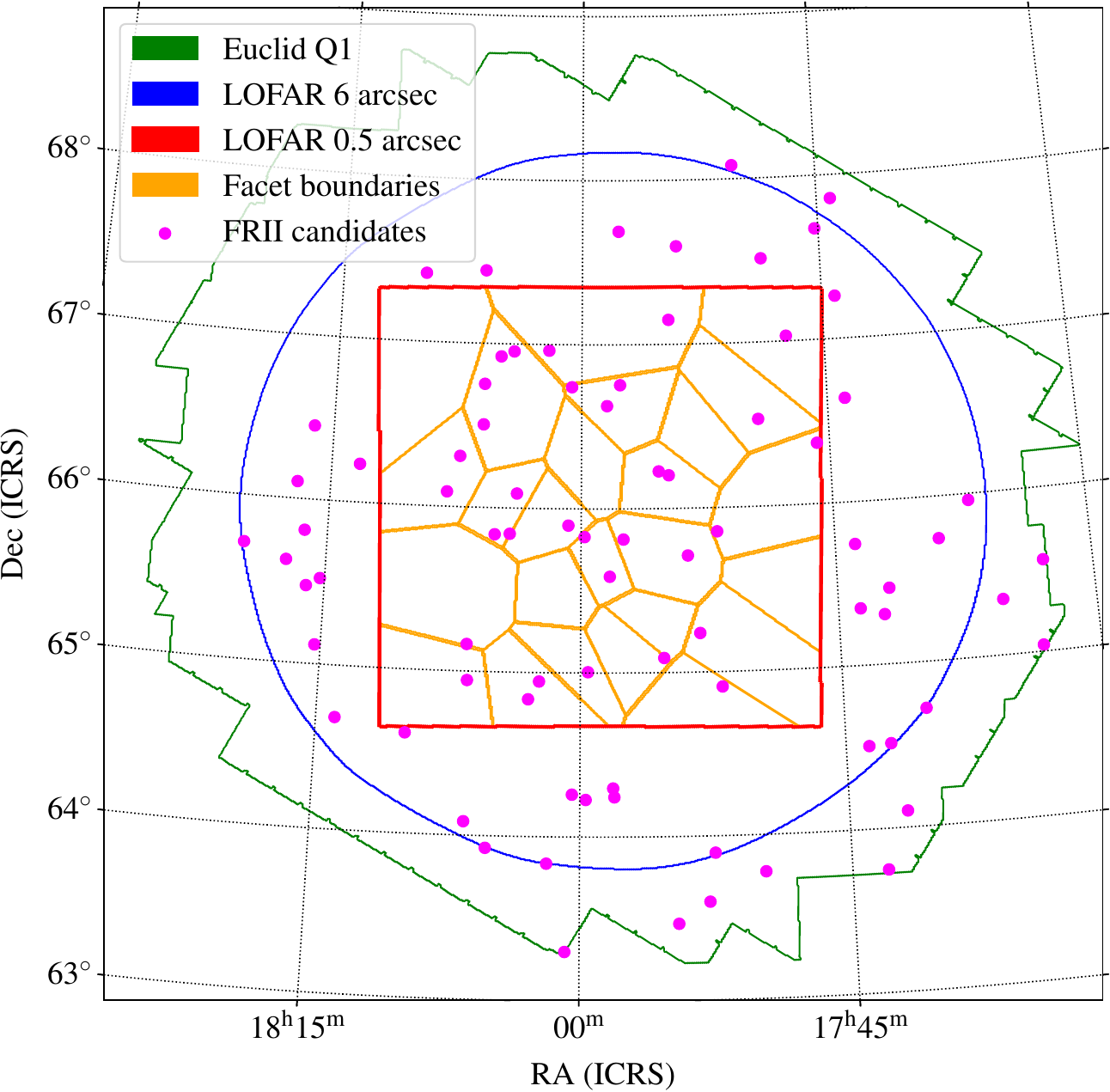}
  \caption{\Euclid\ and LOFAR fields of view used in the
    paper. Lines show the boundary of the area surveyed by 
    \Euclid\ in the Q1 release at \JE\ band \citep{Q1-TP001}, the
    50\% sensitivity boundary of the LOFAR $6''$ data
    \citep{Bondi+24}, and the full area covered by the \mbox{\ang{;;0.5}}
    LOFAR images, together with the internal facet boundaries of those
    images (Bondi et al., in prep.). Also shown are the positions of
    visually selected FRII sources discussed in this paper.}
  \label{fig:fov}
  \end{figure}

\subsection{\Euclid\ }

As noted above, we make use of the Euclid Q1 release in the
EDF-N, which has coverage by high-sensitivity and high-resolution
radio data. Q1 observations cover 22.9\,deg$^2$ of sky in this area
(Fig.\ \ref{fig:fov}) in the visible (\IE) and near-IR (\YE, \JE, and
\HE) bands \citep{Q1-TP001}. We do not use any of the available
ground-based data over the same area as the resolution of \Euclid\ 
is key to our main objective of identification of possible hotspot
counterparts. Visual inspection and photometry use the
background-subtracted mosaics \citep{Q1-TP002,Q1-TP003}, making use of
the fact that \Euclid\ measurements are already in the AB
magnitude system and can easily be converted to flux
density\footnote{In principle, the extreme spectral energy
distributions of some hotspot candidates might affect the photometric
conversion. In practice, the very flat response over the \Euclid\ 
bands means that flux density and effective frequency are not
significantly affected by the sources' spectra, and in what follows we neglect these
effects, which are at the few per cent level.} in Jy.

\subsection{Radio}

The EDF-N is one of the four LOFAR Two-metre Sky Survey (LoTSS) Deep
fields \citep{Best+23} and has been observed with the LOFAR High Band
Antennas (HBA) for around 400 hours, which when fully processed at
6\arcsecond\ resolution will give central rms noise levels of around
10\,$\mu$Jy\,beam$^{-1}$ at 144\,MHz. An initial image at
6\arcsecond\ resolution made from 72 hours of data was presented by
\cite{Bondi+24}, and this has a central rms noise level of
32\,$\mu$Jy\,beam$^{-1}$. Although the full image is much larger and
covers the whole area observed by \Euclid, the inner part of the LOFAR
Dutch-array primary beam, where the sensitivity is greater than 0.5 of
its on-axis value, covers 15.4\,deg$^2$ (Fig.\ \ref{fig:fov}) and we
mainly focus on the area within this coverage in the current paper. An
optical ID catalogue for the 6\arcsecond\ images \citep{Bisigello+25},
based on ground-based data and covering 10\,deg$^2$ of this region, is
also available.

Bondi et al. (in prep.) present the first full-resolution image of
the EDF-N using the international baselines of LOFAR; this image, based
on only 8\,h of LOFAR data at 144\,MHz, has a typical resolution around
\mbox{\ang{;;0.5}} (varying slightly across the field) and a central rms noise
level around 40\,$\mu$Jy\,beam$^{-1}$. Because the primary beam of the
international LOFAR stations is smaller than that of the core or remote
stations, this image covers only 7.1\,deg$^{2}$, around a third of the
\Euclid\ sky coverage and less than half of the coverage of the
6\arcsecond\ image. However, its high resolution is critical to the
accurate location of hotspots and the comparison with the \Euclid\ 
data, as we shall see below. The astrometry of the LOFAR images has
been aligned with the \Euclid\ images by cross-matching point sources.

Finally, we have access to the Very Large Array Sky Survey (VLASS:
\citealt{Lacy+20}) images of the field. VLASS operates at a central
frequency of 3\,GHz and is less sensitive than wide-field LoTSS
data \citep{Shimwell+22} for a typical optically thin synchrotron spectral index (with an
rms noise level typically around 70\,$\mu$Jy\,beam$^{-1}$). Since it is 
missing short baselines, it is not sensitive at all to
extended structures larger than around 30\arcsecond. However, it has the
advantage of having slightly higher resolution than Dutch LOFAR,
around \mbox{\ang{;;2.5}}, and its sensitivity is highest compared to LOFAR
for flat-spectrum source features such as hotspots, while of course it
covers the whole \Euclid\ area. We obtained all
available VLASS `quick look' observations of our target fields, which
span several VLASS observing epochs, and
stacked them to obtain a single image, which was used only as part of the
visual inspection process rather than for photometry.

\subsection{Sample selection and classification}

Sample selection and classification for this project proceeded as follows.
We first visually inspected\footnote{All visual inspection throughout
the paper was done by the lead author.} the whole LOFAR 6\arcsecond\ image for
candidate FRII sources, including barely resolved small double
sources, marking them with \software{ds9}\footnote{\url{https://sites.google.com/cfa.harvard.edu/saoimageds9}} regions. This gave us a total of
130 candidates within the Euclid Q1 sky coverage.

Next, we obtained VLASS images for each of the targets, and inspected
these to reject objects that did not show clear hotspots at the higher
resolution of VLASS, or were otherwise morphologically not consistent
with our selection of FRIIs, excluding objects that were, for example,
clearly lobed FRIs or wide-angle tail sources. This reduced our sample
selection to 80 objects, whose positions (the centres of circular
\software{ds9} regions enclosing the sources) are marked in pink on
Fig.\ \ref{fig:fov}. As a full associated LOFAR catalogue is not yet
available for this sky area, we gave the sources names based on their
approximate radio position of the form JXXX.XX+XX.XX, where the right
ascension and declination are specified in decimal degrees. The full
source list is available in Table A.1\footnote{Available in electronic form only at \url{http://cdsweb.u-strasbg.fr/cgi-bin/qcat?J/A+A/}}.

Finally, we overlaid the available radio data for each source on the
\Euclid\ images at all four bands, and searched for the optical and IR
identification of the radio source (the host galaxy or quasar) and for
plausible optical/IR counterparts of the hotspots, looking
particularly at the peak of the VLASS hotspots. For each source, the
position of the optical identification was noted, and convincing or
possible optical hotspot counterparts were recorded for the two lobes
(denoted for convenience the `north' and `south' lobes irrespective of
source orientation on the sky), where the definition of `convincing'
or `possible' takes account mostly of the offset between the
optical/IR position and the apparent peak of the radio hotspot. We
required a detection in all three near-IR bands in preference to the
optical, since the spectra of hotspots are often expected to be steep,
but we also visually inspected the optical data to obtain the best
resolution for the morphology of optical objects. Optical or IR
objects that were clearly morphologically galaxies on inspection of
the high-resolution optical data, or that were significantly offset
(more than 2--3 arcsec) from the peak of the VLASS radio emission,
were not counted.

After an inspection of all 80 sources in the full area covered by the
6\arcsecond\ LOFAR and \mbox{\ang{;;2.5}} VLASS data, we then
re-inspected the 32 of these targets that were covered by the
high-resolution LOFAR imaging to check the validity of classifications
derived from low-resolution data.

\section{Results}

\subsection{Optical identifications}

Of the 80 sources, 77 had optical or IR hosts in the \Euclid\ images,
while three were hidden by bright foreground objects. FRII
hosts tend either to be quasars or massive galaxies, and so are
readily detectable out to high redshift. In a few cases there was no
detectable central radio component and we picked the most plausible
host galaxy from several candidates based on its position relative to
the lobes, but most (around 90\%) were unambiguous.

We recorded the positions of the host galaxies using \software{ds9}
and cross-matched within 3\arcsecond\ with, in order of preference,
the DESI DR1 spectroscopic catalogue \citep{DESI25}, the photometric
redshifts for DESI Legacy Survey sources estimated for the whole DESI
sky by \cite{Duncan22} provided that the quality cut of that work was
satisfied, and the optical identification catalogue of
\cite{Bisigello+25}, which contains photometric redshift estimates
derived using \software{cigale} (we use the Bayesian estimator of
redshift) for $z<5$. We also obtained a few spectroscopic redshifts
from the literature by searching the NASA Extragalactic Database (NED)
for objects within 3\arcsecond\ of the host positions, which were
preferred over photometric redshifts if only the latter were
available, adopting what appeared to be the best-quality spectrum in
each case. In total 64 objects had a redshift estimate from either
spectroscopy or photometry, with the redshifts lying in the range $0.3
\la z \la 3.0$. For the objects with redshifts, we calculated radio
luminosity at rest-frame 144\,MHz on the assumption of $\alpha = 0.7$.
Properties of all 80 sources are presented in Table A.1, which also
indicates the sources missing an optical ID (no optical ID position)
and those missing a redshift estimate.

\subsection{Fraction of candidate counterparts in low-resolution images}

Three objects were dropped at the visual inspection stage in which we
overlaid radio and optical data, since the position of their host
galaxies ruled out an FRII interpretation. Out of 77 confirmed FRIIs,
we found 11 IR hotspot counterpart candidates that we classified as
high-confidence (Table A.1), by which we mean that
there was an offset of less than 1\arcsecond\ between the optical
position and the peak of the VLASS image, and a further 14 where there
is a possible counterpart that is less clearly aligned. The IR
emission from four hotspots was hidden by a bright foreground object
(stars or in one case the Cat's Eye planetary nebula, NGC\,6543). Thus
the statistics are 11/73 (15\%) and 25/73 (34\%) for high-confidence
and all possible counterparts respectively.

We can make an estimate of the chance coincidence rate expected by
using the Q1 source catalogue over all the tiles in the
EDF-N area. This
contains 3\,307\,894 objects that are detected\footnote{Here we define
`detection' as a signal-to-noise ratio greater than 3 for each of \YE, \JE,
and \HE, which is probably conservative, since we do not
accept any object that does not have clear counterparts by eye in all
three near-IR bands.} over the three IR bands over the 22.9\,deg$^2$ of
coverage. If we assume that an IR source has to lie within
1\arcsecond\ 
(the VLASS pixel size) of the VLASS peak hotspot position to be
identified as a hotspot, and we consider two possible counterparts per
source, then we expect $2 \times 73 \times \pi \times 1.0^2 \times
3\,307\,894/(22.9 \times 3600^2) \approx 5$ spurious `counterparts', with the
number obviously rising steeply as the requirement for positional
cross-matching is relaxed. This estimate does not account for sources
that show obvious galaxy morphology in the optical images, which would
have been rejected in our visual inspection, but it is
clear that spurious sources could account for a significant fraction
of the total here. High-resolution radio images are needed to verify a good
positional alignment between candidate optical and IR hotspots and their
radio counterparts.

\subsection{Candidates in high-resolution images}

Of the FRII targets, 32 lie in the 7.1\,deg$^2$ of high-resolution
(\mbox{\ang{;;0.5}}) LOFAR coverage. Of these, three (out of the four
mentioned previously) are sources whose hotspots are hidden in the 
  \Euclid\ data by foreground objects. Five had been flagged from the
lower-resolution inspection as high-confidence hotspot detections and
five as possible detections. The remaining 19 sources with
high-resolution LOFAR observations did not have hotspot detections in
overlays with the 6\arcsecond\ images.

\begin{figure*}
  \includegraphics[height=0.30\linewidth]{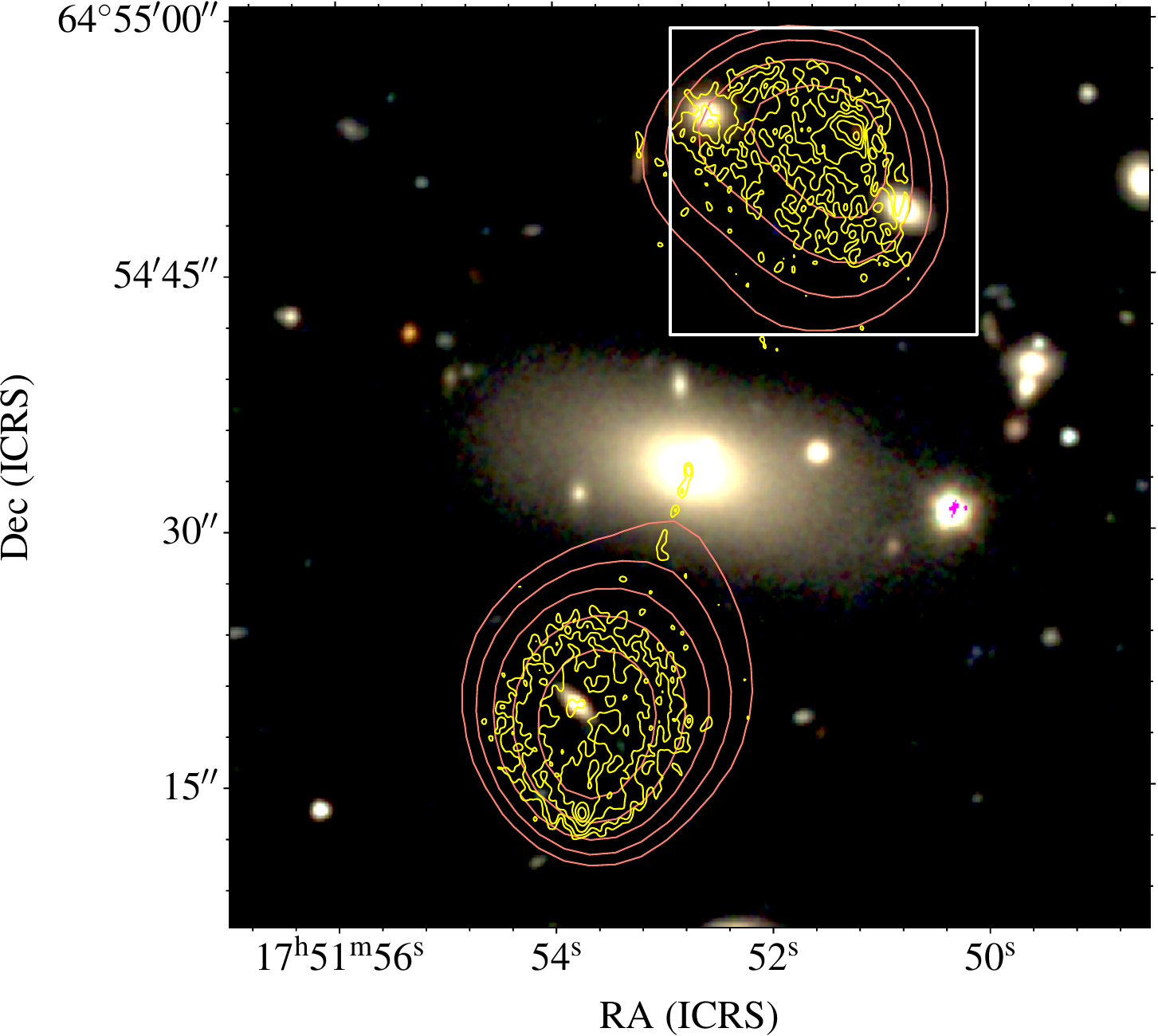}
  \includegraphics[height=0.30\linewidth]{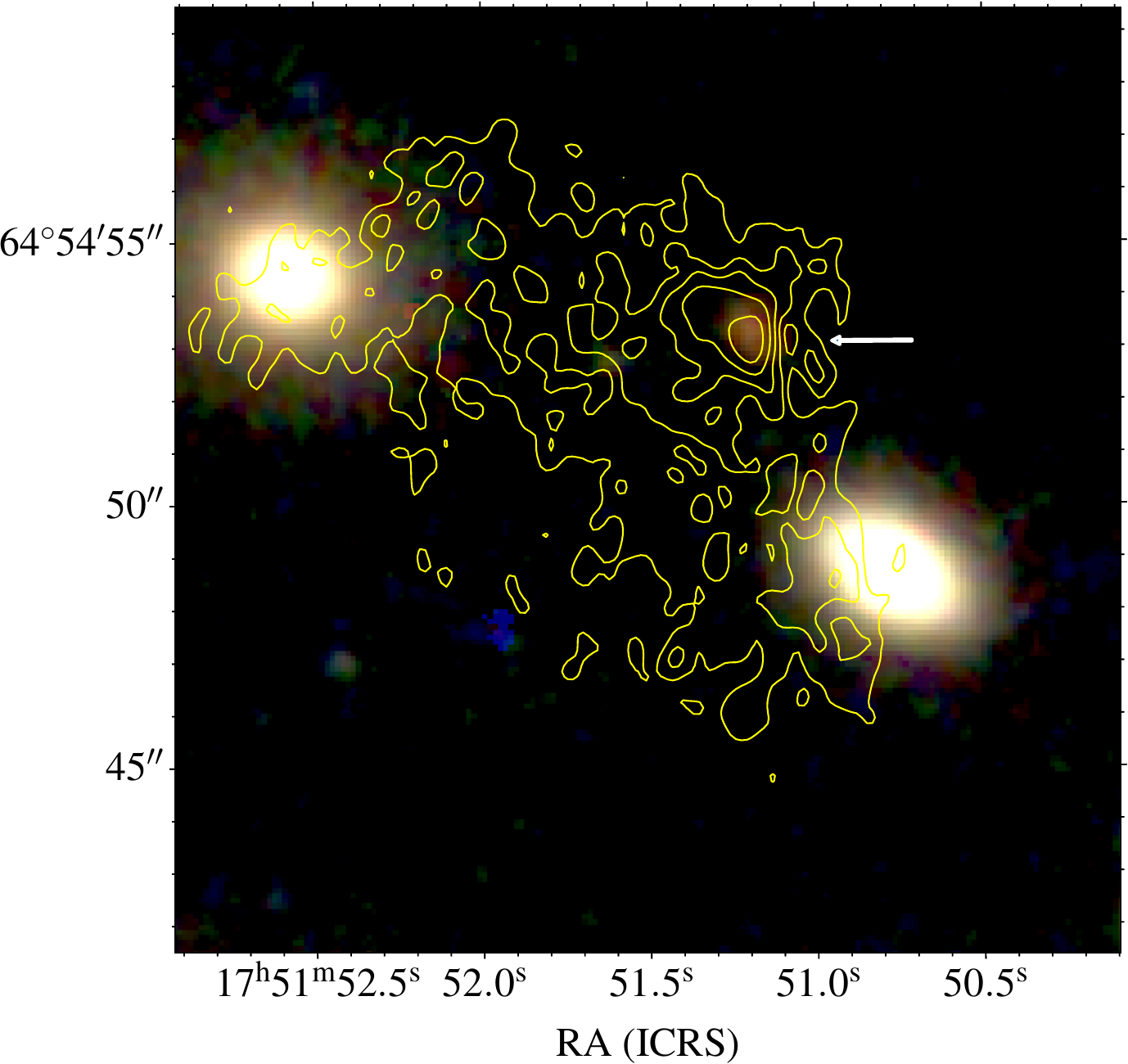}
  \includegraphics[height=0.30\linewidth]{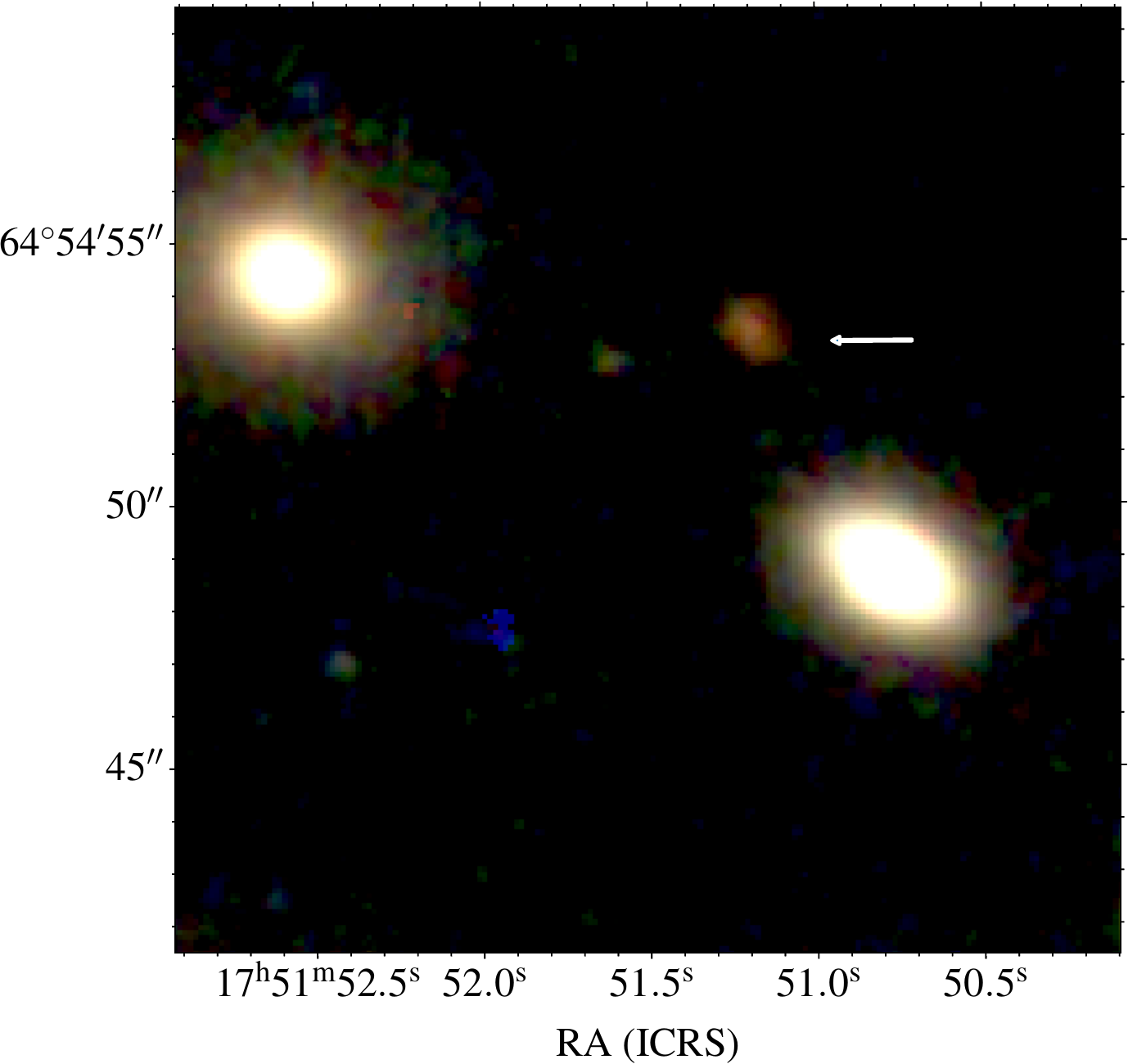}\\
  \includegraphics[height=0.30\linewidth]{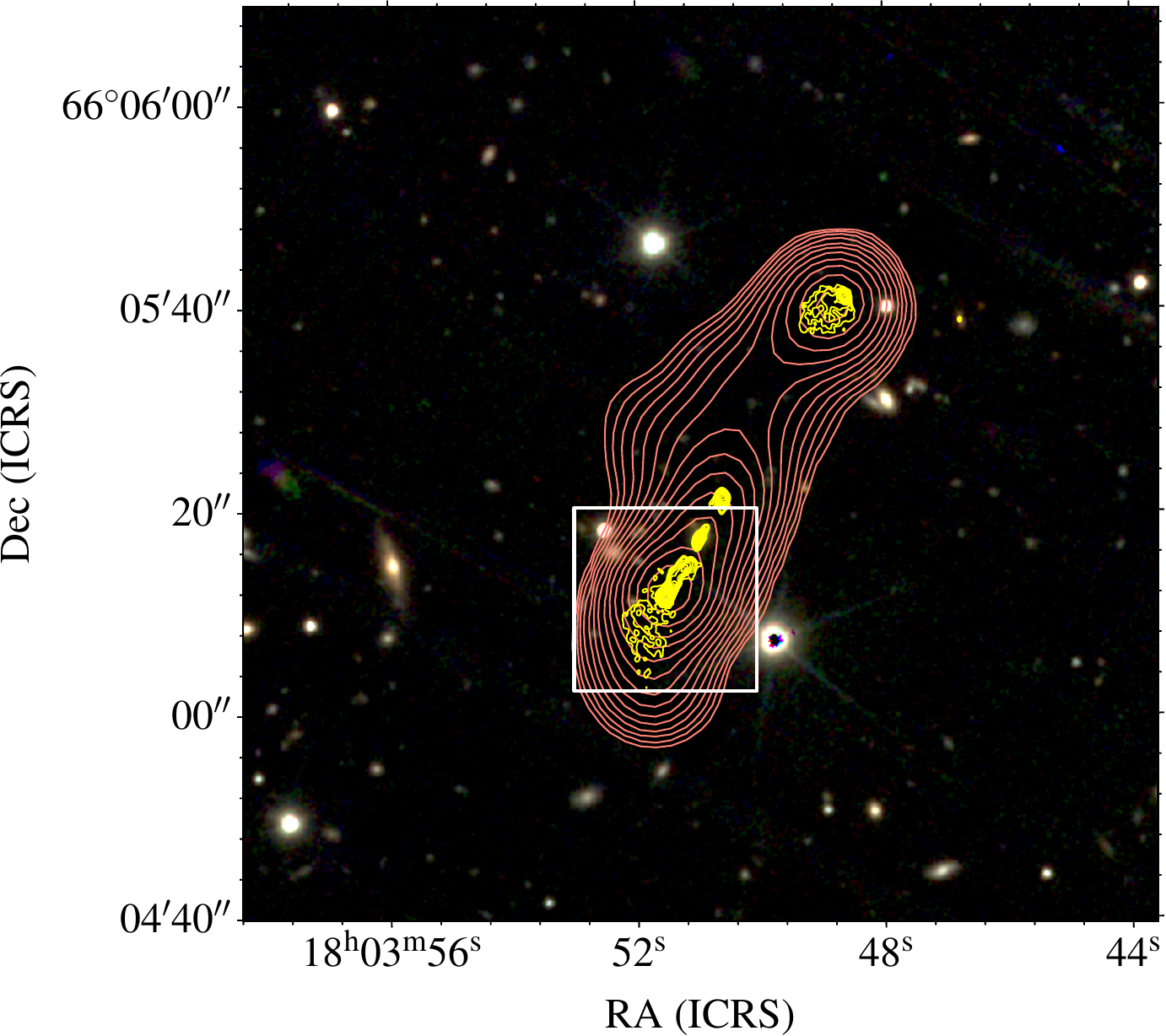}
  \includegraphics[height=0.30\linewidth]{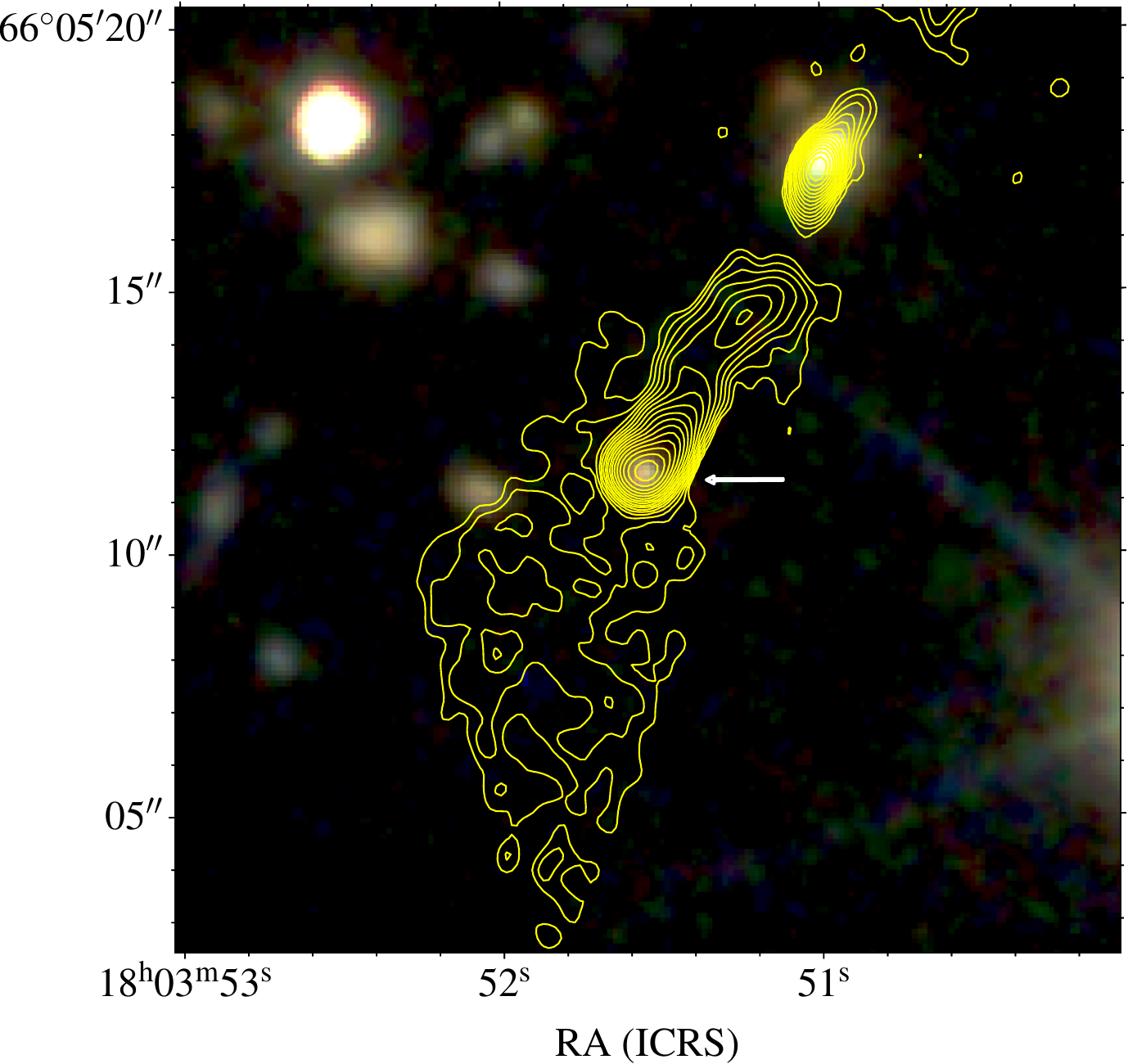}
  \includegraphics[height=0.30\linewidth]{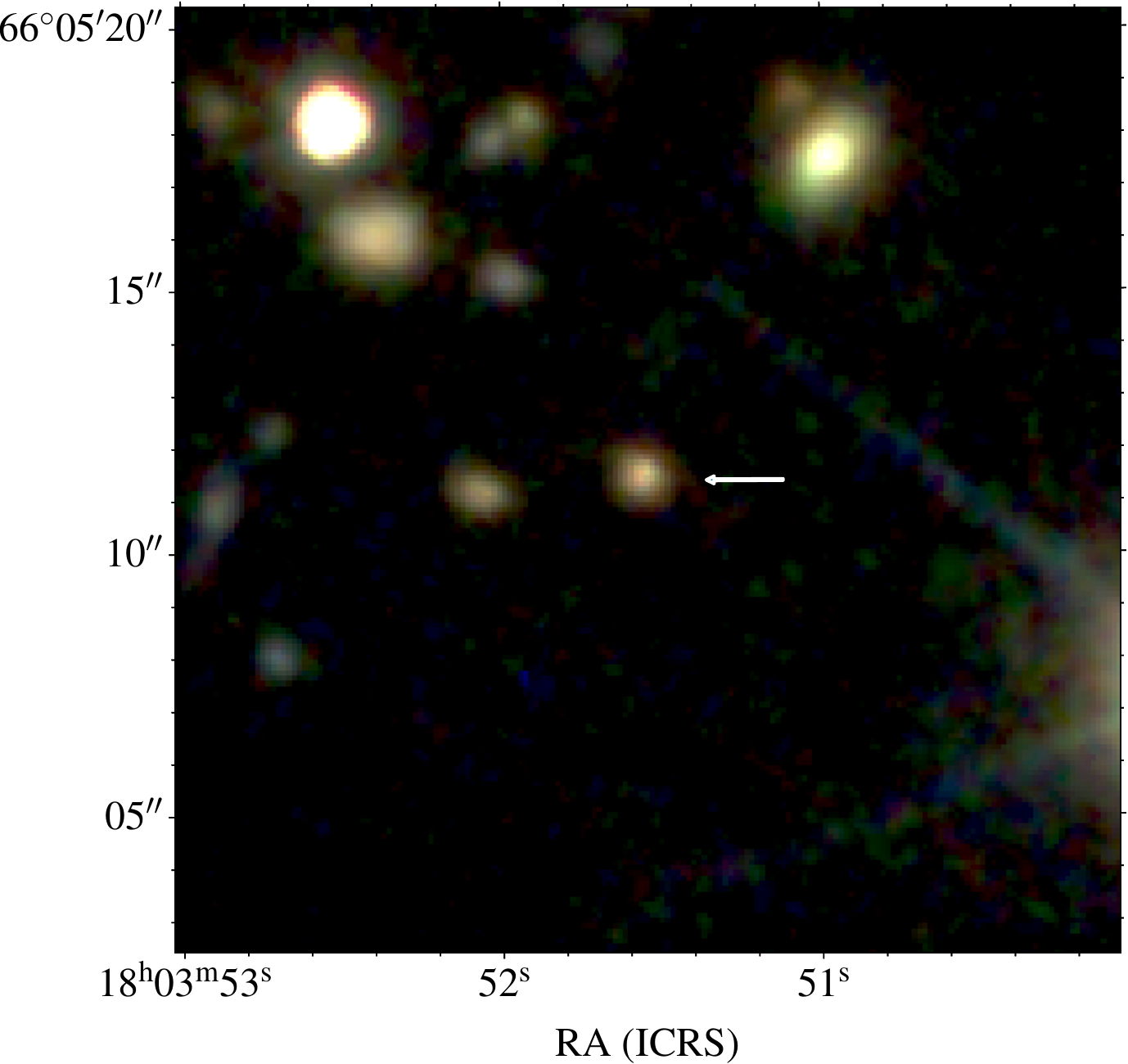}
  \caption{Images of the FRII targets with confirmed close matches
    between radio and near-IR hotspots. The top row shows
    J267.966+64.91 (6C B175142.4+645504 at $z=0.294$; \Euclid\ detects a counterpart to
    the northern compact hotspot). The bottom row shows J270.961+66.09
    (6C B180350.0+660455 at $z=1.607$; \Euclid\ detects a counterpart to the southern hotspot of the inner double of a
    restarting source, while the host galaxy is also visible to the
    northwest in the hotspot image). The colours show a
    false-colour representation, with a square root transfer function,
    of the \Euclid\ IR bands. The contour levels are logarithmic; yellow contours show the radio
    emission at \mbox{\ang{;;0.5}} resolution, while pink contours show the
    $6''$
    LOFAR map. Left column shows the full radio source, with zoom
    region marked. The middle column
    shows a zoom in on the detected hotspot, with only high-resolution
    radio contours and with an arrow indicating
    the position of the optical source. The right column shows the same as
    the middle column, but with the radio contours removed.}
  \label{fig:detections}
\end{figure*}

\begin{figure*}
  \includegraphics[height=0.30\linewidth]{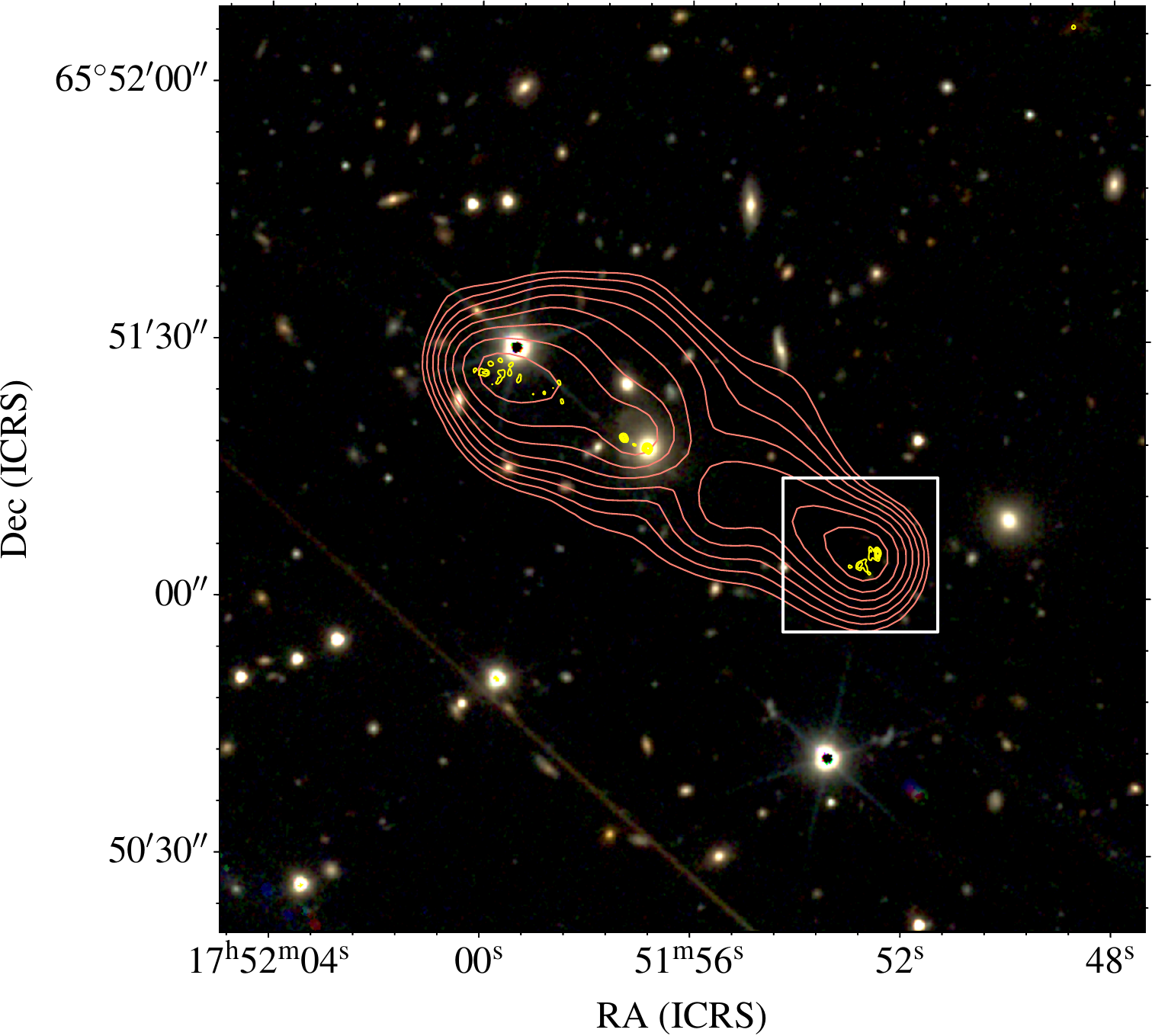}
  \includegraphics[height=0.30\linewidth]{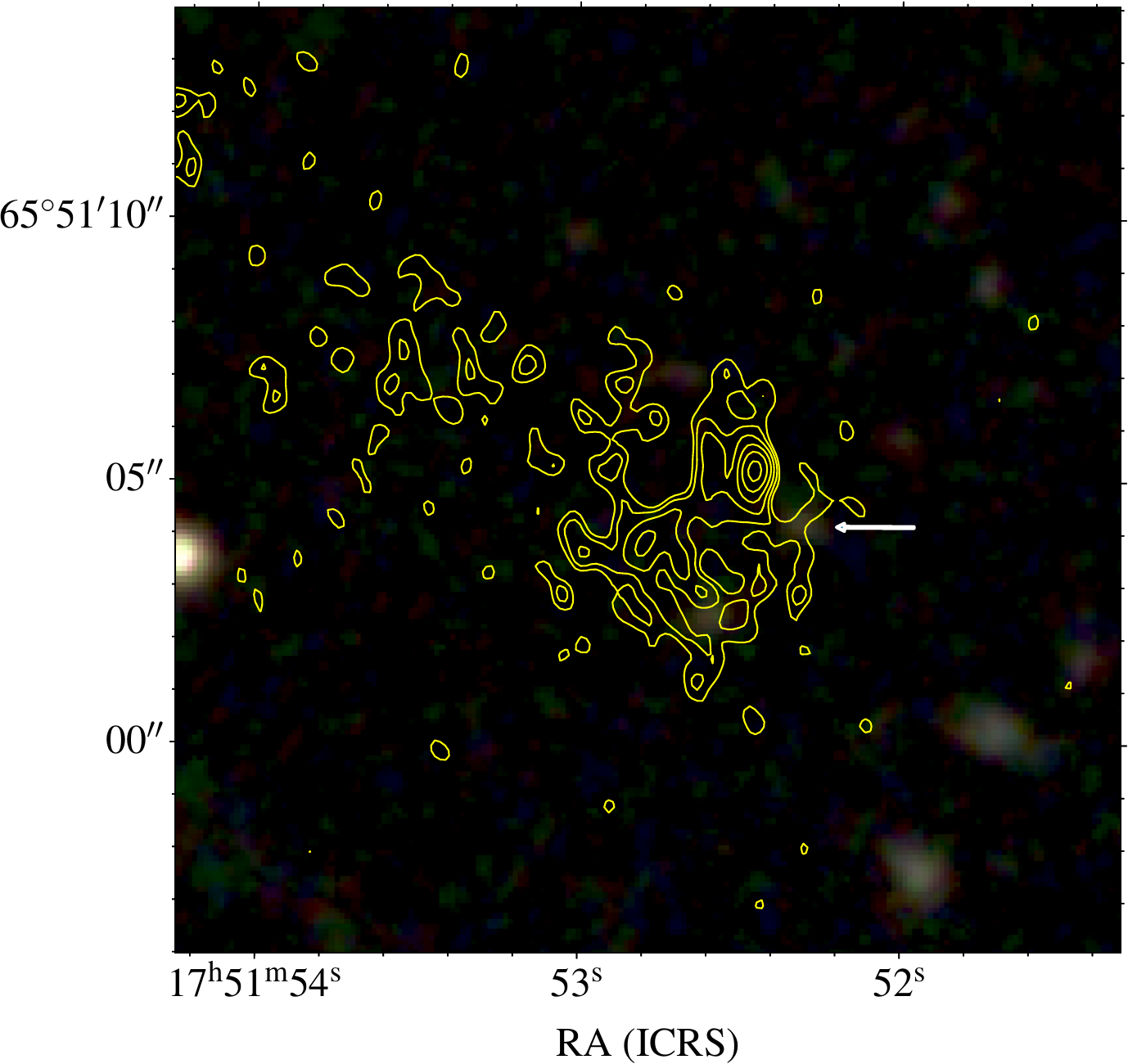}
  \includegraphics[height=0.30\linewidth]{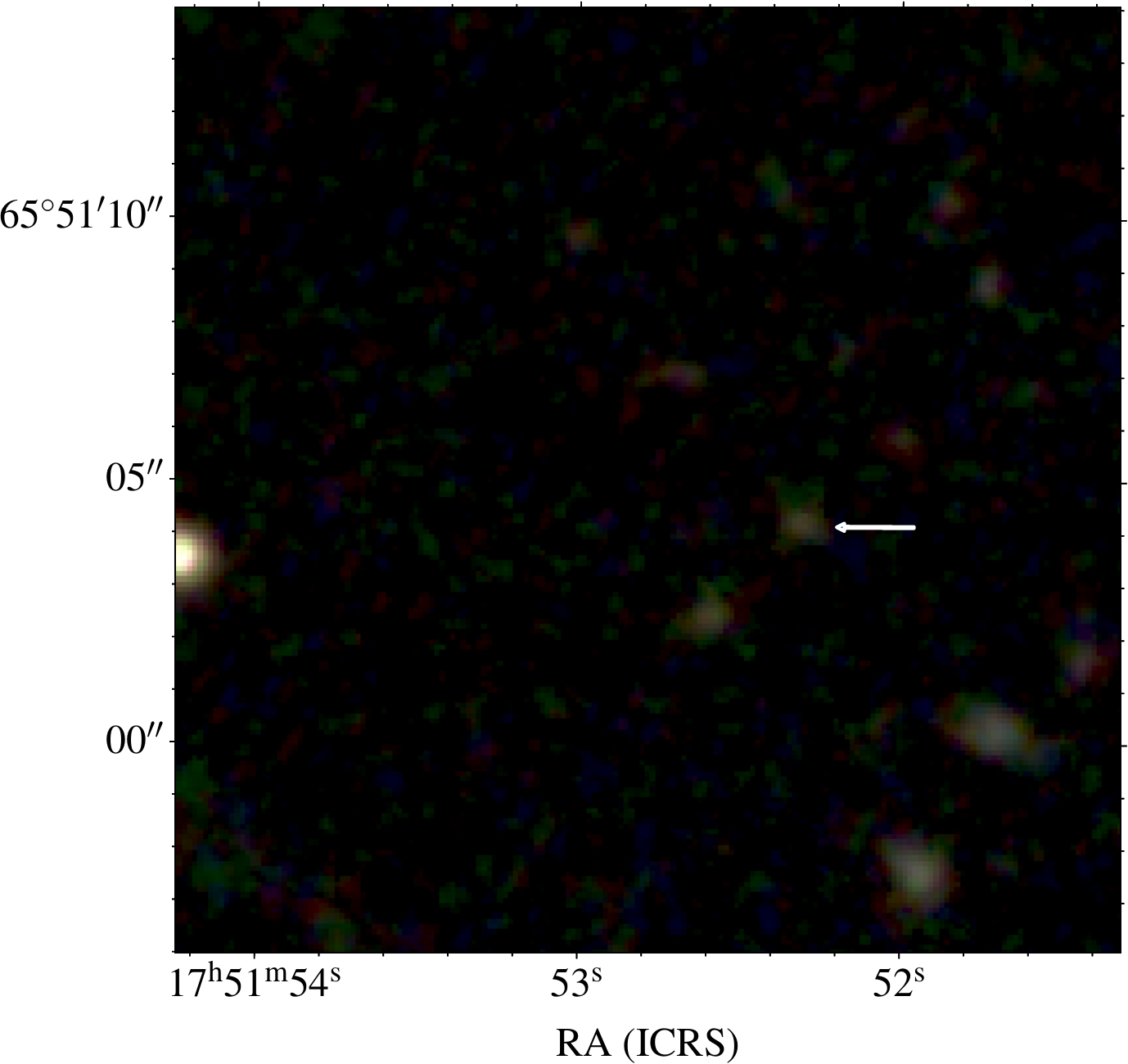}\\
  \includegraphics[height=0.30\linewidth]{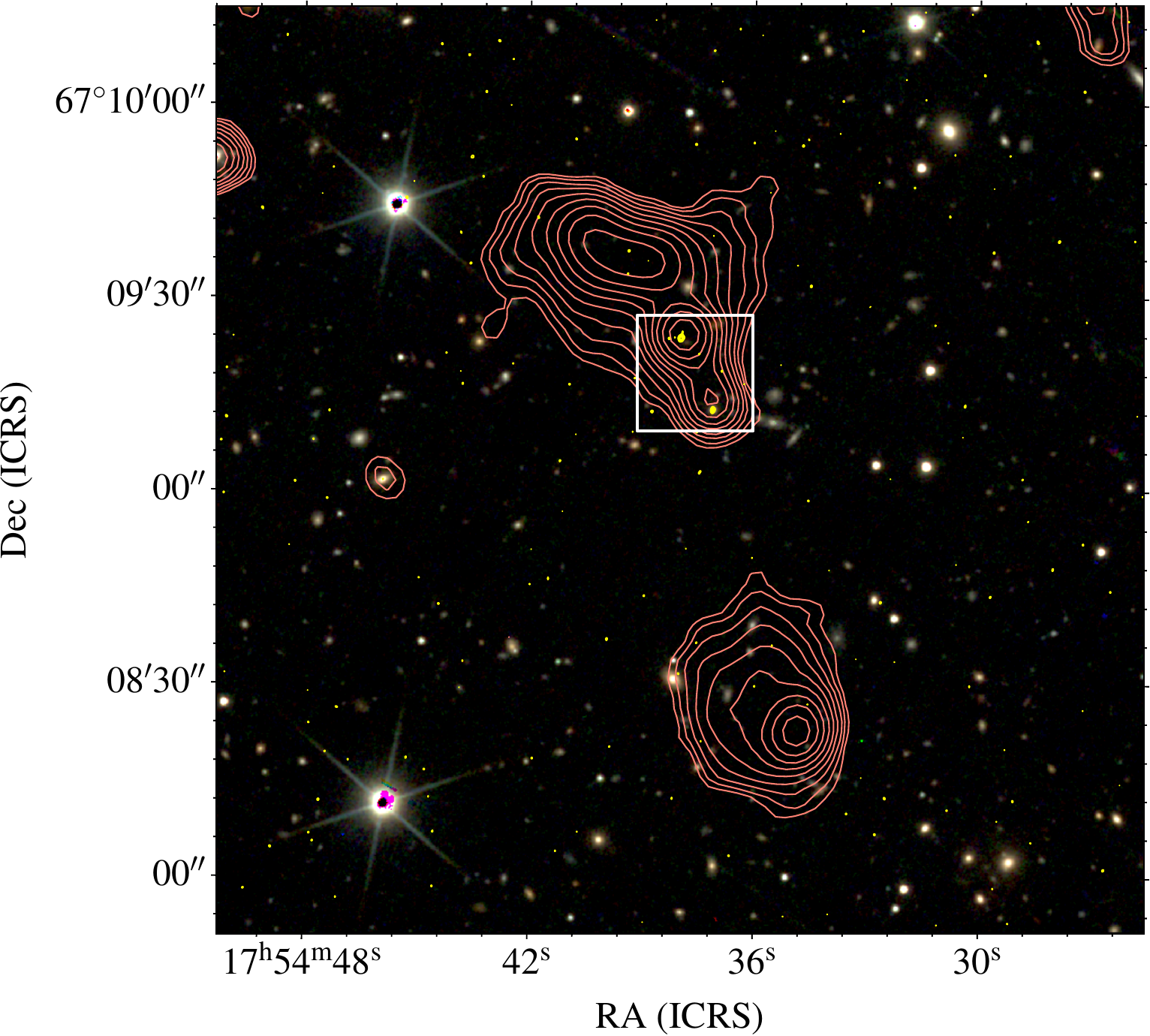}
  \includegraphics[height=0.30\linewidth]{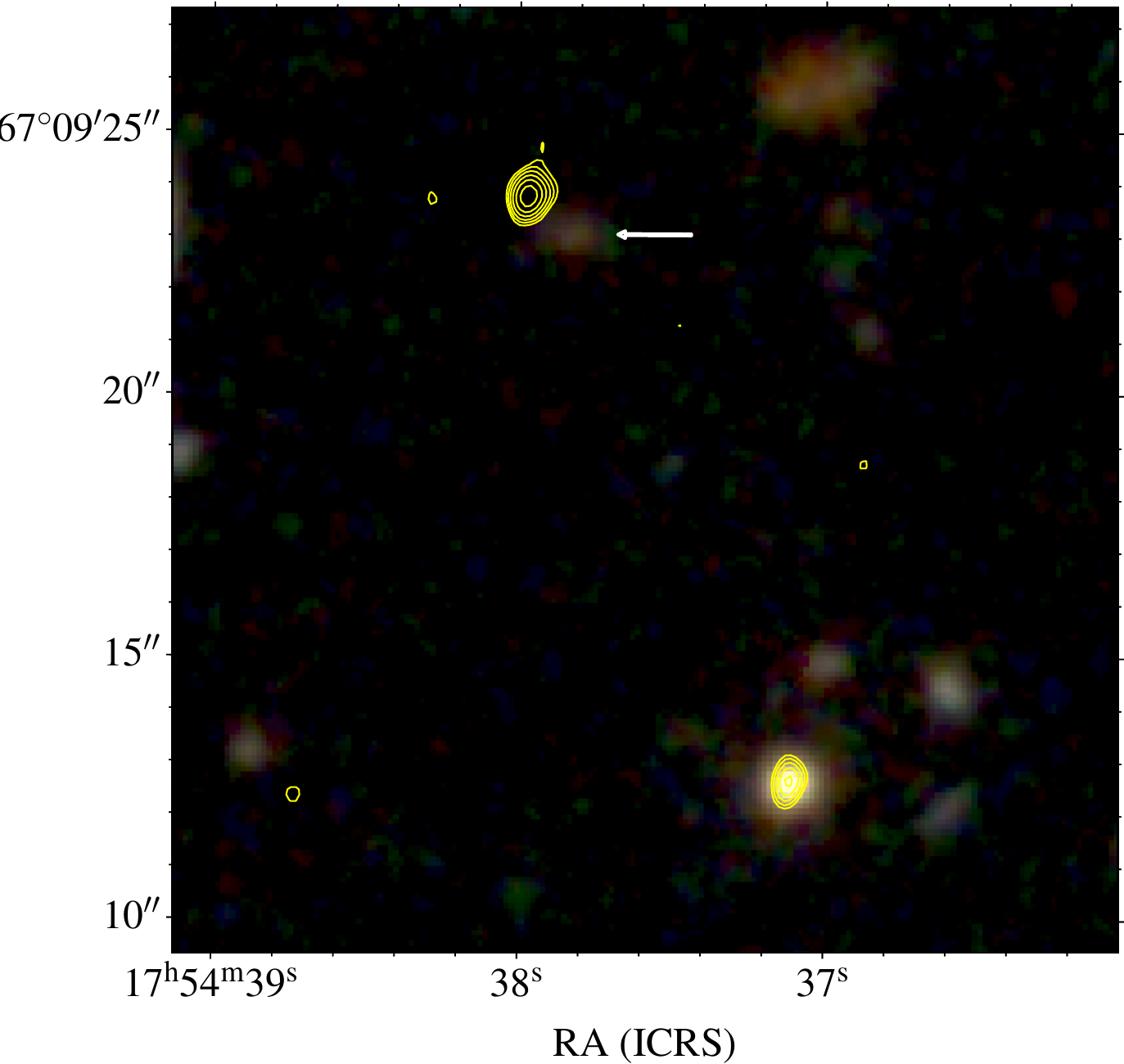}
  \includegraphics[height=0.30\linewidth]{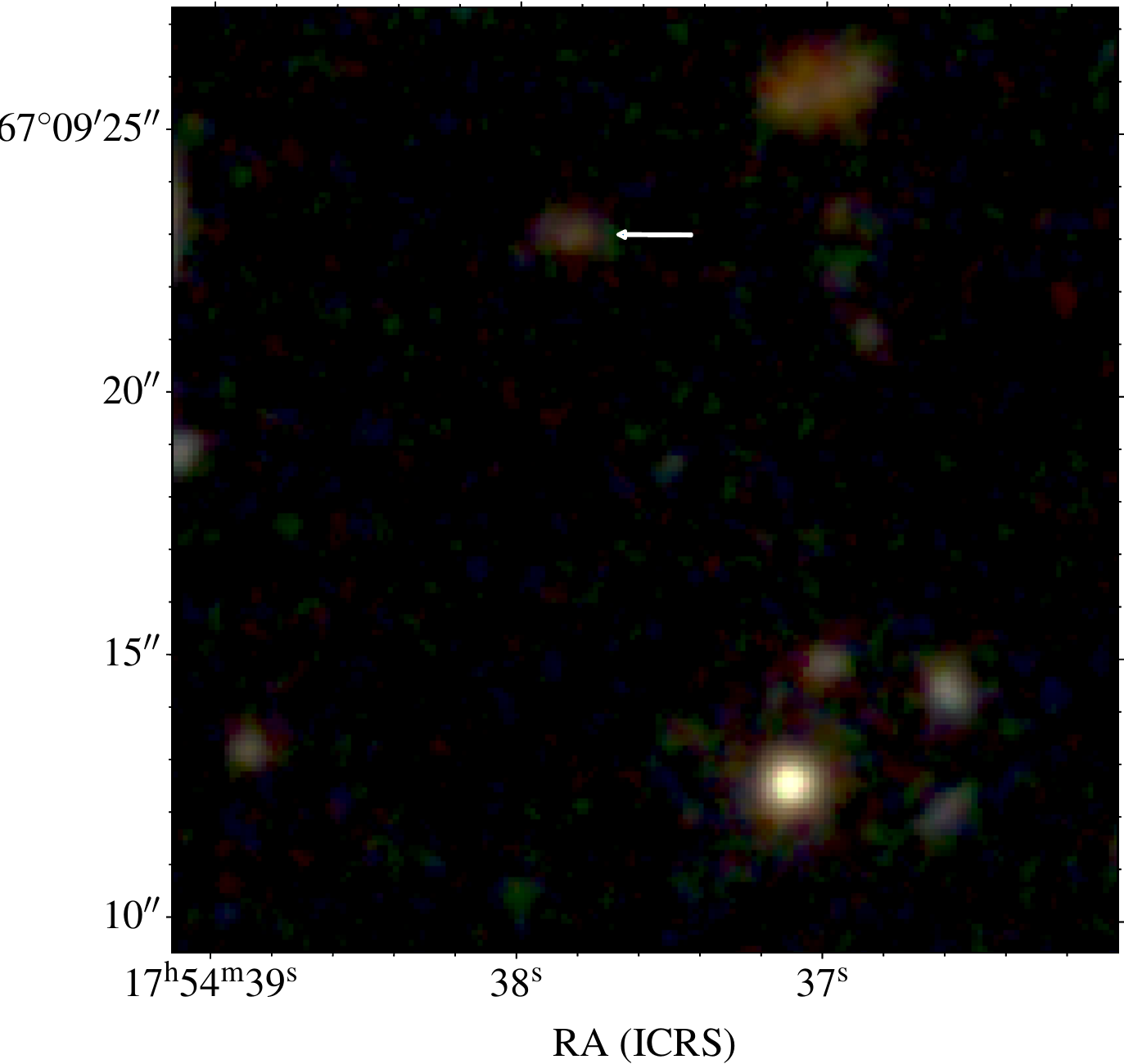}\\
  \includegraphics[height=0.30\linewidth]{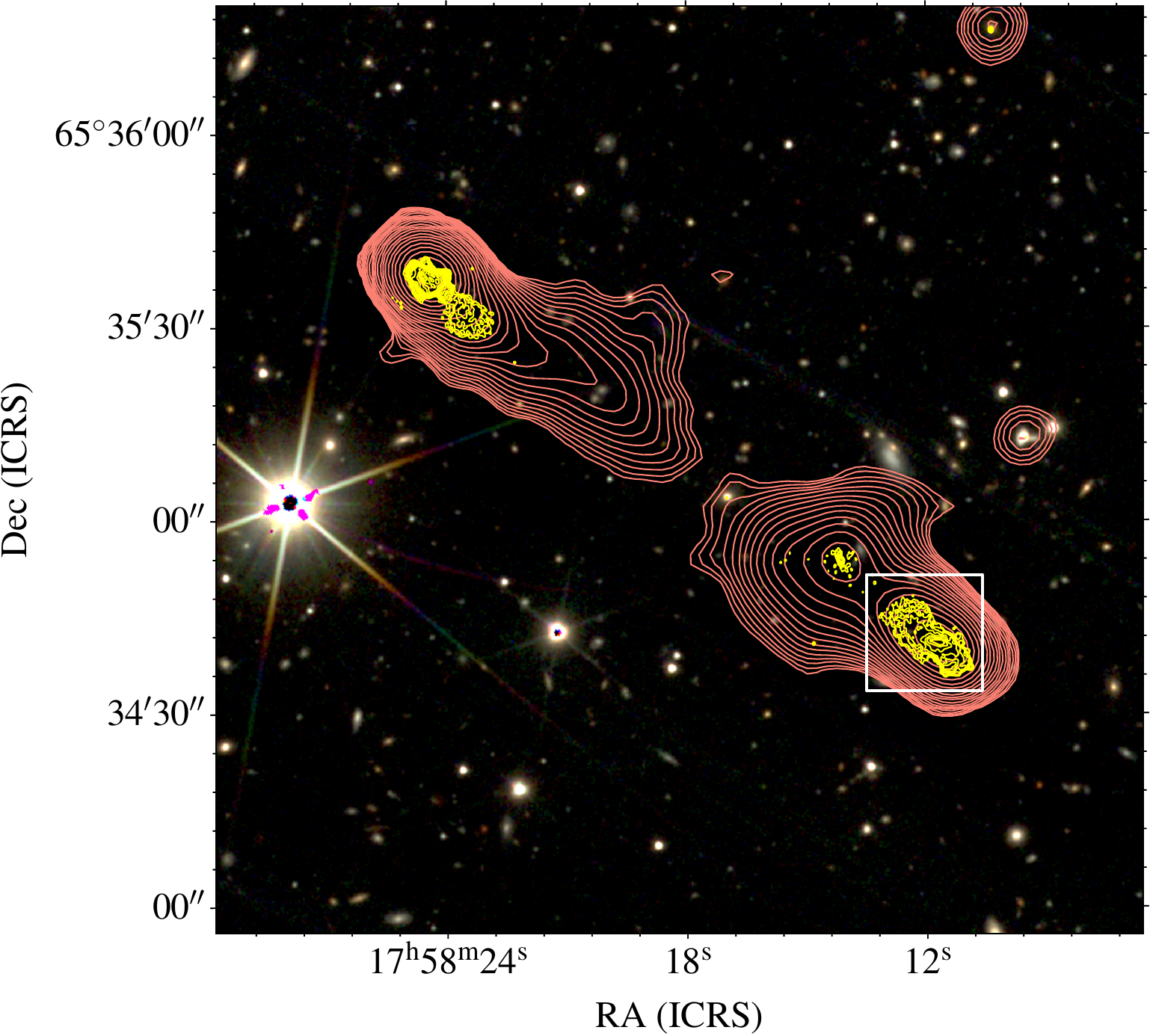}
  \includegraphics[height=0.30\linewidth]{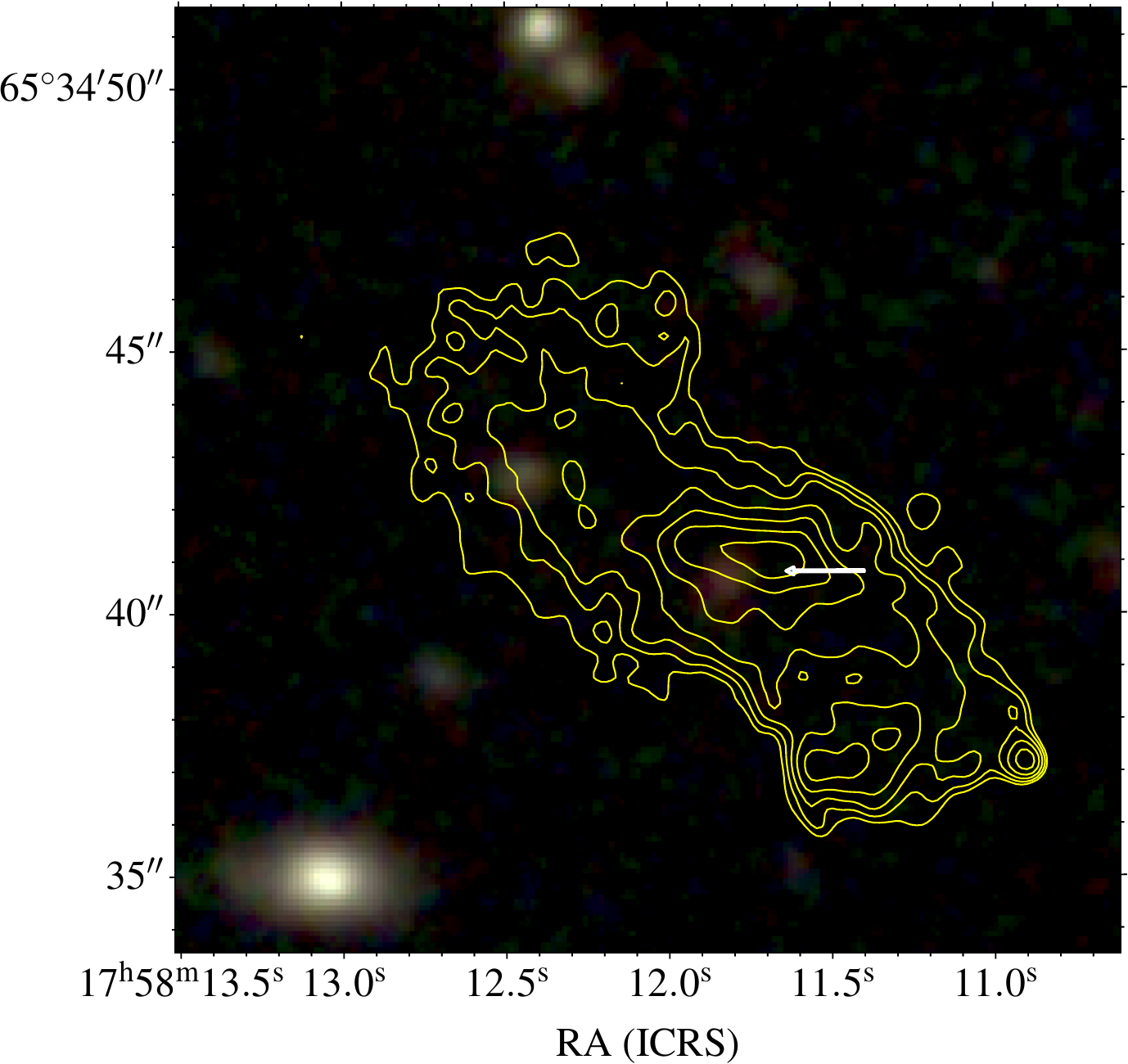}
  \includegraphics[height=0.30\linewidth]{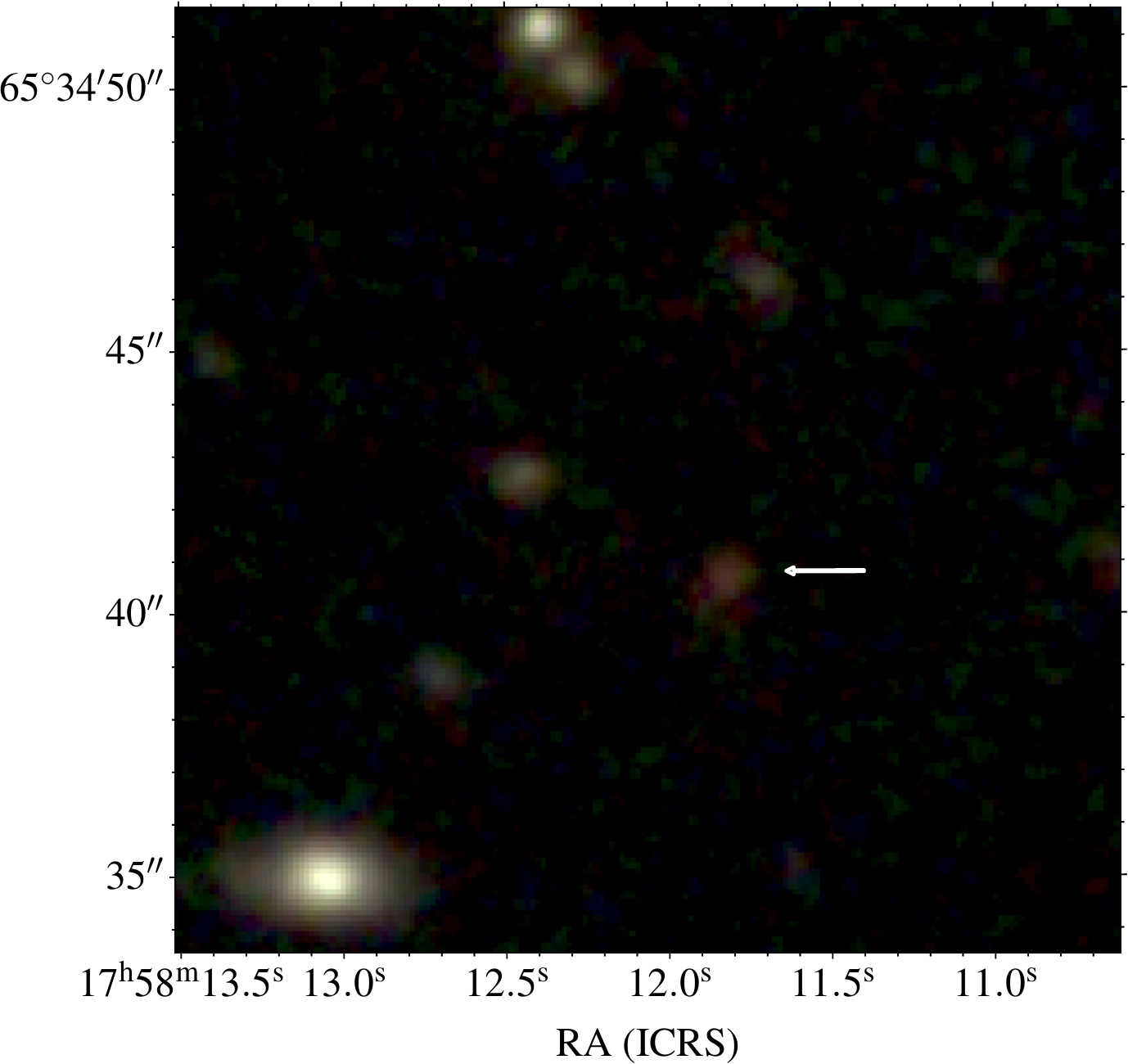}
  \caption{Images of the FRII targets where a `good confidence'
    IR counterpart from VLASS images was not confirmed as a
    positional match to compact features in
    high-resolution images. From top to bottom, rows show J267.986+65.85
    (IR counterpart is not aligned with the only compact
    component in the hotspot region),
    J268.657+67.15 (IR counterpart is offset from jet knot) and
    J269.576+65.59 (IR counterpart is within the brightest radio
    region, but not coincident with a radio peak). Colours and contours as in
    Fig.\ \ref{fig:detections}, but the zoom is into the candidate
    hotspot from $6''$ imaging.}
  \label{fig:notconfirmed}
\end{figure*}

On inspection of the high-resolution data, we confirmed two out of five of
the `high-confidence' detections, J267.966+64.91 and J270.961+66.09,
as having sub-arcsecond positional agreement between a compact hotspot
in the radio and a clearly detected IR source. These two confirmed
detections are shown as radio and \Euclid\ overlays in
Fig.\ \ref{fig:detections}. Figure \ref{fig:notconfirmed} shows the
same overlays for the three out of five sources where we do not confirm an IR
hotspot detection. One of these three, J269.576+65.59, remains a
possible IR hotspot candidate, since the IR counterparts are in a region
of bright complex radio emission, but they are not associated with a
discrete radio source. The other two `high-confidence' detections from
the 6\arcsecond\ imaging, J267.986+65.85 and J268.657+67.15, show
significant positional offsets between the peak of the radio emission
and the \Euclid\ source and are probably not true detections. None of
the five `possible' detections was confirmed with the high-resolution
data (although one remains possible) nor was any previously unseen
candidate found. Thus the high-confidence detection fraction stands at
$2/29 \approx 7$\% (Table A.1), and the possible
fraction at $4/29$. With the sub-arcsec positional alignment of the IR
and radio counterparts in our two best candidates it is very unlikely
that these are chance alignments with unrelated objects. On the same
calculation as carried out above, we expect $\ll 1$ spurious
counterparts given the much smaller radio beam and pixel size.

\section{Discussion}

\subsection{Interpretation of the detected sources}

We measured flux density for the two confirmed high-resolution radio
and \Euclid\ hotspots by integrating over a circular region centred on
the near-IR detection. Flux densities obtained for the different
hotspots are given in Table \ref{tab:flux}.

\begin{table*}
  \begin{center}
    \caption{Flux densities of the confirmed hotspots}
    \label{tab:flux}
  \begin{tabular}{lrrrrr}
    \hline
    \hline
    Source&\multicolumn{5}{c}{Flux density ($\mu$Jy) at}\\
    &144 MHz&\YE&\JE&\HE&\IE\\
    J267.966+64.91&$(23 \pm 2) \times 10^3$&$4.6 \pm 0.1$&$2.3 \pm
    0.1$&$0.6 \pm 0.1$&$0.44 \pm 0.02$\\
    J270.961+66.09&$(126 \pm 13) \times 10^3$&$7.6 \pm 0.1$&$4.7 \pm
    0.1$&$2.8\pm0.1$&$0.37 \pm 0.02$\\
    \hline
  \end{tabular}
  \end{center}
\end{table*}

\begin{figure}
  \includegraphics[width=\linewidth]{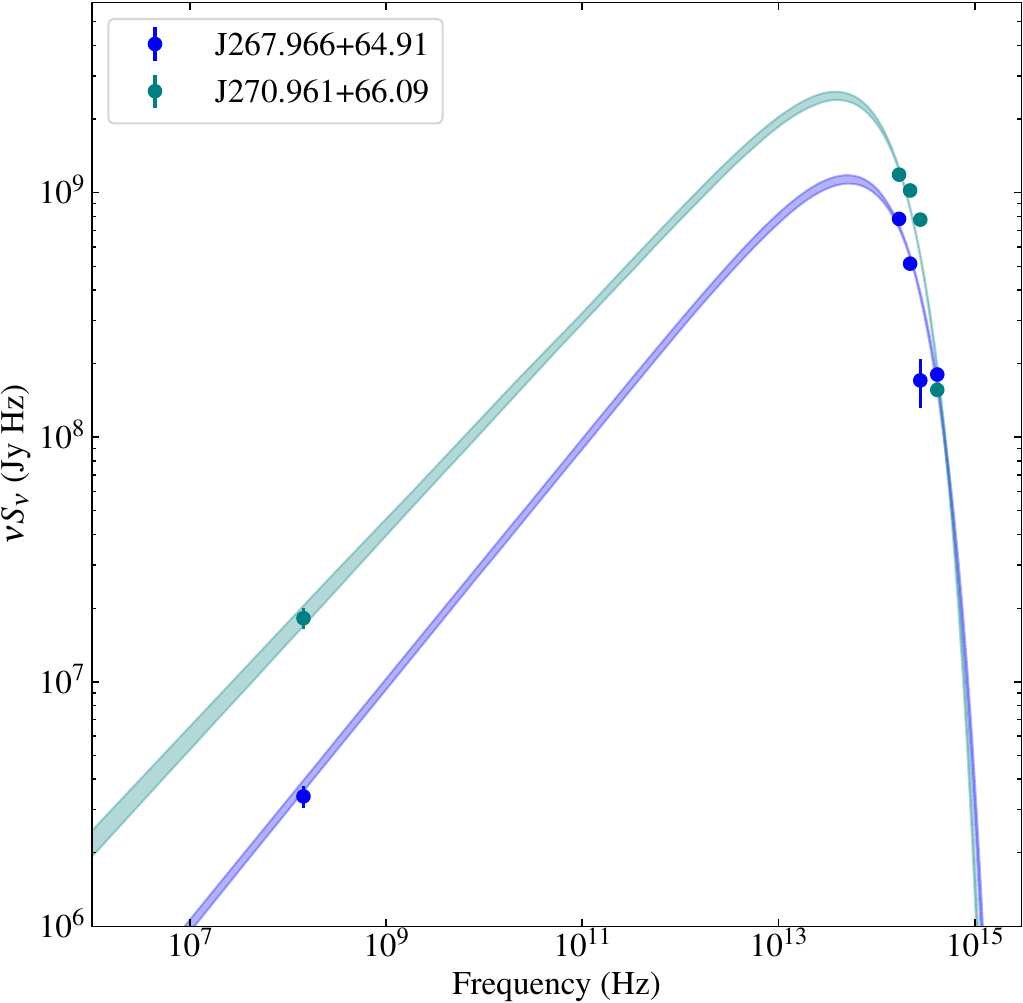}
  \caption{Spectral energy distributions of the two best hotspot
    candidates with the 68\% credible interval range of samples
      from the MCMC fits overplotted. Error bars (68\% confidence,
    statistical foreground error only) on the data points are shown but are smaller than
    the symbols in some cases. See the text for details of the models
    shown.}
  \label{fig:hsfit}
  \end{figure}

We fitted the spectral energy distributions of the hotspots with the
model used by previous work such as that of \cite{Meisenheimer+97},
which is the one-zone continuous-injection electron energy spectrum
expected from particle acceleration at a strong shock
\citep{Heavens+Meisenheimer87}. Given the lack of radio photometry
other than from LOFAR, we used a simple power-law electron energy
spectrum with a fixed power-law index, a fixed break and a high-energy
cutoff,
\[
N(\gamma) = \begin{cases} 0, & \gamma < \gamma_\mathrm{M},\\
  N_0 \gamma^{-p}, & \gamma_\mathrm{M} < \gamma < \gamma_\mathrm{B},\\
  (\gamma_\mathrm{B} N_0)\, \gamma^{-(p+1)}, & \gamma_\mathrm{B} < \gamma < \gamma_\mathrm{C},\\
  0, & \gamma \ge \gamma_\mathrm{C}\\
\end{cases}
\]
where $\gamma$ is the electron Lorentz factor [$\gamma =
  E/(m_\mathrm{e}c^2)$], $\gamma_\mathrm{M} = 1$, $\gamma_\mathrm{B}$
and $\gamma_\mathrm{C}$ are the Lorentz factors of the break and the
cutoff respectively, which are varied to fit the data, $p$ is the `injection index' of accelerated particles, which depends on the shock physics, and $N_0$ is
the electron number normalization derived from the radio flux density and magnetic field strength.
This corresponds in synchrotron radiation terms to a steepening from
$\alpha = (p-1)/2$ to $\alpha = p/2$, followed by an exponential cutoff.
For a strong shock, we expect $p=2$ \citep{Bell78}, which corresponds to a low-frequency spectral index $\alpha=0.5$, but weaker shocks might be expected to give higher values of $p$.
  
Fitting an electron energy distribution to an observed synchrotron spectrum requires an assumption about the magnetic field strength $B$ in the radiating electrons. For many years it has been conventional to make use of the equipartition assumption, in which the energy density in the field and the particles is the same; historically this was selected because it also comes close to minimizing the energy density requirements to produce a given synchrotron emissivity \citep{Burbidge56}. Inverse-Compton observations of well-studied hotspots in the X-ray \citep[e.g.][]{Harris+94,Hardcastle+04} have shown that the magnetic field strength is in fact typically a little below the equipartition value in all objects studied. Field strengths much lower than equipartition would give rise to very bright inverse-Compton emission in the X-ray and even potentially the optical bands, but this is not observed, although the sample size for which the X-ray observations are possible is small.
  
We therefore used the electron energy distribution
  presented above to estimate indicative magnetic field strengths on
the assumption of equipartition from a spherical emission region with
an angular radius of 1\arcsecond, on the conservative assumption of no
non-radiating particles. Synchrotron spectra were calculated using the
code of \cite{Hardcastle+98-2}, implemented in the \software{pysynch}
package\footnote{\url{https://github.com/mhardcastle/pysynch}}. Fitting
  treated overall normalization, $p$, $\gamma_\mathrm{C}$ and
  $\gamma_\mathrm{B}$ as free parameters, using \software{emcee} to
  carry out a Markov-Chain Monte Carlo (MCMC) sampling the posterior
  of a $\chi^2$-based likelihood space with a uniform prior for $p$
  ($2 \le p \le 3$), and Jeffreys priors for $\gamma_\mathrm{C}$ and
  $\gamma_\mathrm{B}$ ($5 \le \log_{10}(\gamma) \le 7$). Results are
  tabulated in Table \ref{tab:fits}, where the quoted error bars are
  marginalized over all other parameters, and plotted in
  Fig.\ \ref{fig:hsfit} as samples from the posterior.

\begin{table*}
  \begin{center}
    \caption{Bayesian estimates of derived electron energy spectrum parameters for the two hotspots}
    \label{tab:fits}
  \begin{tabular}{lrrrrr}
    \hline
    \hline
    Source&$p$&$\log_{10}(\gamma_\mathrm{C})$&$\log_{10}(\gamma_\mathrm{B})$&$B_\mathrm{eq}$ (nT)\\
    J267.966+64.91&$2.01 \pm 0.01$&$6.14 \pm 0.01$&$6.5 \pm 0.3$&$2.31 \pm 0.05$\\
    J270.961+66.09&$2.13 \pm 0.02$&$6.034 \pm 0.01$&$6.4 \pm 0.3$&$6.5 \pm 0.2$\\
    \hline
  \end{tabular}
  \end{center}
\end{table*}
  
By selection, given that these are faint radio hotspots detected in
the optical/IR, they are `low-loss' hotspots by the definition of
\cite{Meisenheimer+97}, in the sense that the high-energy break and
cutoff appear in the synchrotron spectrum only close to the
infrared/optical regime. In fact, for both fits, the preferred
  model does not include a break at all (that is, the break energy is
  constrained to be above the cutoff energy). J267.966+64.91 was best
  fitted with $p = 2.0$ and $\gamma_\mathrm{C} \approx 1.4 \times
  10^6$, while the fit to J270.961+66.09 preferred a slightly steeper $p=2.13$ and
  $\gamma_\mathrm{C} \approx 1.1 \times 10^6$. Because the
synchrotron emission peak frequency of an electron varies as
$\gamma^2B$, where $B$ is the magnetic field strength, these energies
would be slightly increased if the magnetic field strength were below
the value corresponding to equipartition with the radiating leptons
alone by a small factor. It is worth noting that the low values
  of $p$ in the fits -- which are still consistent with the fits to
  some of the sources studied by e.g. \cite{Meisenheimer+97} -- arise
  because high $p$ values would cause the fits to underpredict the
  emission in the \YE\ band. The two-point spectral indices between
  144 MHz and \YE\ are 0.59 and 0.67 for J267.966+64.91 and
  J270.961+66.09 respectively, which give hard upper limits ($p <
  2.18$, $p < 2.34$) on the injection indices, and in fact because of
  the finite width of the synchrotron kernel in frequency space the
  fitted values of $p$ have to be substantially lower than this to
  allow the fit to reproduce the clearly visible steep cutoff ($\alpha
  \gg 1$) in infrared frequencies. In this sample, where we start from
  comparatively faint radio sources and are thus limited by the depth
  of the \Euclid\ data, we may expect a bias towards low values of $p$
  in detected sources relative to earlier studies of bright 3C sources
  with dedicated optical observations.
  
An important caveat is that one-zone models such as the one
  described above are
unlikely to represent the complexity of particle acceleration in real
hotspots. We know both from observation
\citep[e.g.][]{Hardcastle+97,Carilli+99,Tingay+08,Hardcastle+16,Orienti+20}
and from simulation \citep[e.g.][]{Horton+23} that there is likely to be
spatial structure internal to the hotspots that we cannot resolve (at
these distances) with either our radio or optical observations; there
may even be non-negligible temporal structure to hotspot light curves
that observations taken at different times would be integrating over.
Nevertheless, these one-zone models, treating the hotspot as a
homogeneous sphere, do provide perhaps surprisingly
good fits to the overall integrated hotspot spectral energy densities
for both synchrotron and, where present, inverse-Compton emission
\citep[e.g][]{Meisenheimer+89,Meisenheimer+97,Harris+94,Hardcastle+01}
which suggests that they are not completely inaccurate. A plausible
conclusion is that the small-scale structure represents the location
of instantaneous particle acceleration, but that the bulk of the
emission from a hotspot is coming from more homogeneous regions where
point-to-point magnetic field variation, while still present, does not strongly
affect calculations of the mean field strength or the inverse-Compton
emissivity \citep[cf.][]{Hardcastle13}.

With that caveat, we draw two key conclusions from the fits of
Fig.\ \ref{fig:hsfit}. Firstly, the fits with these simple one-zone
models are reasonably good in the \Euclid\ bands; the spectral energy
distributions are consistent with a synchrotron model of the type that
has been used for other optical hotspots. Secondly, if we take the IR
hotspot detections at face value and assume a field strength at or
moderately below equipartition, these hotspots are currently
accelerating electrons to TeV energies ($\gamma_\mathrm{C}
m_\mathrm{e}c^2$, where the electron mass-energy $m_\mathrm{e}c^2$ is
0.5\,MeV), but, crucially, not above those energies in great numbers,
since the fits show that the IR-through-optical spectra are consistent
with seeing the exponential cutoff in the synchrotron spectrum
associated with the high-energy cutoff in the electron energy
spectrum. It is interesting to note that none of the other compact
radio features of these sources is an optical or IR source. In both
cases it is the brightest radio hotspot in the source that has the
optical counterpart, but scaling the \Euclid\ photometry by the ratio
of the radio fluxes, we would have expected to detect other compact
features in the sources if they had the same intrinsic spectral energy
distribution with $p \approx 2$. Thus we can conclude that those other
hotspots or jet knots are currently not accelerating electrons to such
high energies, or have a very different spectral energy distribution
from the two detected compact features in some other way, perhaps with
much higher $p$ values, or that the optical emission from them is
suppressed, for example by low magnetic fields or Doppler boosting out
of our line of sight.

Finally, we note that the source J270.961+66.09 (6C\,B180350.0+660455: Fig.\ \ref{fig:detections}) is
remarkable in its own right in two ways; it is to our knowledge both
the first optical hotspot detected in the inner double of a restarting
source\footnote{Restarting or `double-double' sources
\citep{Schoenmakers+00} are a class of objects in which two aligned pairs of
lobes are seen, indicating that the jet has switched off for a period
before restarting. The inner hotspots in these objects are expected to
be the current locations of jet termination, though outer hotspots may
also be seen if the jet has not been switched off long enough for them to fade, or if light-travel time effects are important.}, and the highest-redshift optical synchrotron counterpart known, with a spectroscopically confirmed host galaxy redshift of $z\approx1.6$. Samples of optical synchrotron counterparts to bright radio galaxies such as the 3CRR sample have tended to lie at much lower redshifts. This illustrates the value of the combination of \Euclid\ and LOFAR, which collectively can probe a much broader range of source properties than has hitherto been possible.

\subsection{Hotspot detection statistics and particle acceleration}

\begin{figure*}
  \includegraphics[width=0.49\linewidth]{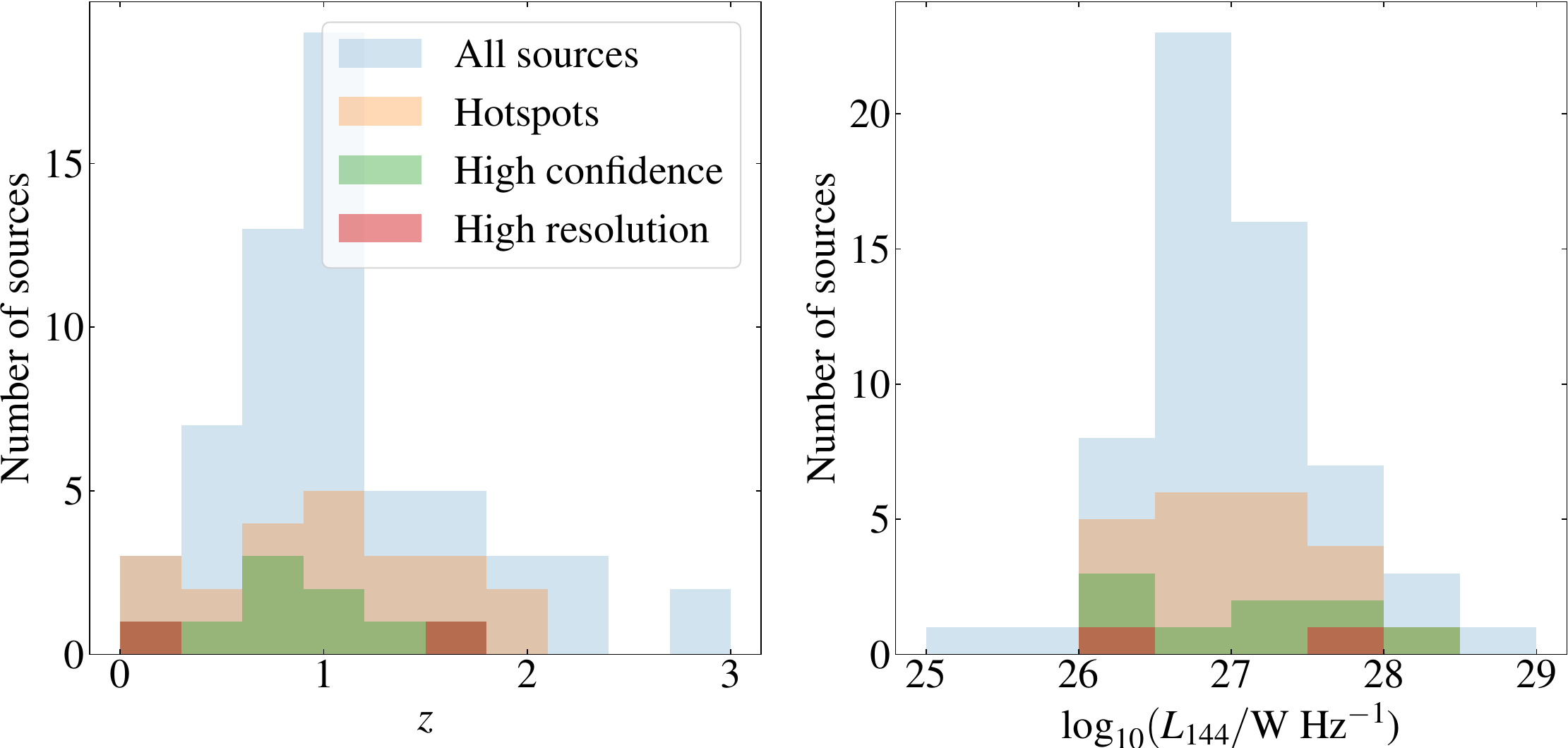}
  \hskip 0.01\linewidth
  \includegraphics[width=0.49\linewidth]{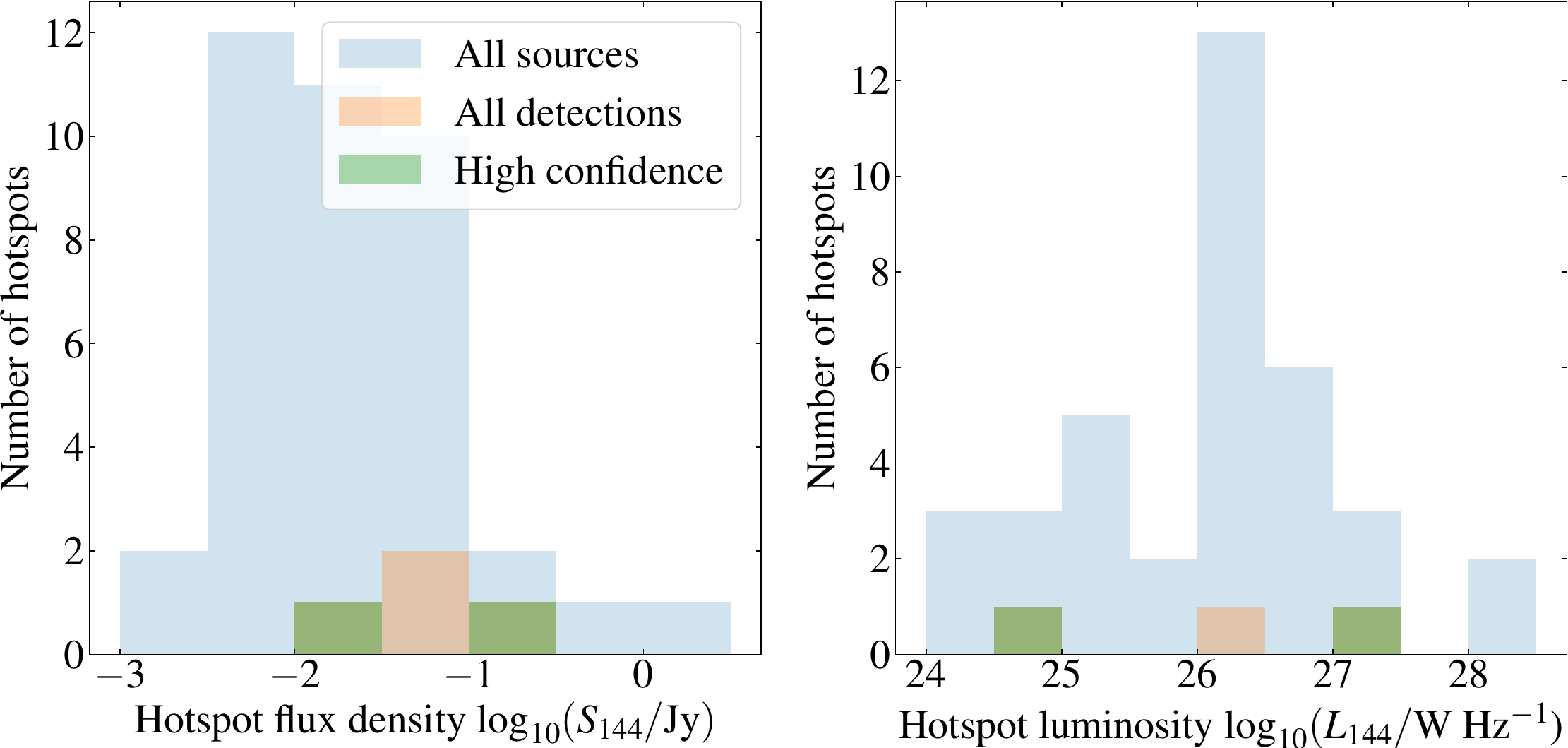}
  \caption{Distributions of hotspot properties. Left two panels: the distribution of the FRII sample in redshift and radio
    luminosity, with all possible, high-confidence and high-resolution
    confirmed hotspots overplotted. There is no evidence for any
    influence of luminosity or redshift on the detectability of
    optical hotspots in this sample. Right two panels: the
    distribution of flux density and radio luminosity for detected and
    non-detected hotspots seen in the high-resolution radio data. Only
  three detected hotspots are shown in the rightmost panel, since one
  possible detection does not have a redshift.}
  \label{fig:hshist}
  \end{figure*}

Plotting histograms of the distribution of detected hotspots as a
function of radio source properties, we find that there is no apparent
relationship between the statistics of possible detections of optical
hotspots and the physical properties of the radio sources hosting them
(Fig. \ref{fig:hshist}); the null hypothesis that the sources with
candidate hotspots are distributed like the parent population in
luminosity or redshift cannot be rejected on a two-sample
Kolmogorov--Smirnov test. In earlier work \citep{Hardcastle+04} the
`low-loss' hotspots, where optical or X-ray synchrotron emission was
prominent, were only found at the lower-luminosity end of the FRII
population, with $26 < \logten(L_{144}/\mathrm{W\,Hz}^{-1}) < 28$, but
since this is the range where almost all of our FRII sources are
found, no strong conclusions can be drawn. It is not known whether
there are any trends with source age (which is related to both
luminosity and physical size), due to the small sample available.
However, it is worth noting that our original selection criteria,
which included a resolved FRII-like appearance in 6\arcsecond\ images,
only includes physically large sources with sizes $\ga 100$\,kpc, and
we are missing a population of smaller objects that might behave
differently. The optical hotspot properties of physically very small
objects are hard to study since they are obscured by the starlight of
the host galaxy, but there will be a range of source properties that
can be probed by combining \Euclid\ and high-resolution LOFAR imaging
that we have not explored here.

A possible hypothesis, discussed by \cite{Hardcastle+04}, is that the
most luminous hotspots, with the strongest magnetic fields, do not
allow the acceleration of particles to the highest energies. In this
context we made radio flux density measurements of all the compact
hotspots found in the 32 FRIIs in the high-resolution area (39
distinct hotspots in total), and find that while the flux densities
and luminosities of the few detected or possibly detected hotspots in
the high-resolution region span a wide range (Fig.\ \ref{fig:hshist}),
it is noteworthy that the brightest and most luminous radio hotspots in our
sample (both in the $z=2.1$ object J270.591+64.95) are not detected in
the near-IR or optical, despite being more than an order of magnitude
higher in flux density than our brightest detection in J270.961+66.09.
For this object, by considering all possible spectra of the form above with $p=2$
that do not imply detection by \Euclid, we can place an upper
limit on the high-energy cutoff or break of the one-zone electron
population, ether $\gamma_\mathrm{C} < 3 \times 10^5$ or $\gamma_\mathrm{B} \ll 3 \times
10^5$, or both, making it clearly different from the detected objects. This
object also has an integrated radio luminosity ($5 \times 10^{28}$\,W\,Hz$^{-1}$) that lies above the range in which FRIIs tend to show
optical hotspots in previous studies.

\section{Conclusions}

\subsection{Implications of the current work}

Our results show that \Euclid\ data provide a robust basis for
identifying IR and optical counterparts of hotspots in powerful FRII
radio galaxies. While the detection fraction is modest, we constrain
it to between around 7\% and 15\% of all visually confirmed FRII
sources. Our most convincing detections required the use of
high-resolution radio data, specifically in our case the
\mbox{\ang{;;0.5}} 144-MHz LOFAR data, to obtain a good positional
match with the optical/IR counterpart. High-resolution radio data are
essential to mitigate confusion from unrelated faint optical sources
and to ensure reliable hotspot identification.

We have seen that in our best cases the broad-band
radio-through-optical spectral energy distributions are very
consistent with the types of models that have been fitted to hotspots
in earlier work, and they require the presence of electrons with roughly
TeV energies (for magnetic field strengths close to equipartition) to
explain the optical synchrotron emission. Conversely, in the brightest
non-detections, we can either rule out the presence of such high-energy
electrons or must assume either very sub-equipartition magnetic field
  strengths, very different electron energy distributions, or some combination of the two. This confirms the physical picture that emerged from
earlier studies of samples of (typically much more luminous) 3C
objects such as those of \cite{Meisenheimer+97} and
\cite{Hardcastle+04}: particle acceleration in hotspots is not
homogeneous. It remains possible that only low-luminosity hotspots
with low magnetic field strengths (and hence low radiative losses) can
accelerate particles effectively to these high energies, but the
detection of only some of the hotspots within our best candidate
sources points to a more complex picture in which other factors are
also at play. With the detection of a high-$z$ IR hotspot candidate,
we have also established the potential of searching for evolution in
hotspot properties as a function of cosmic time.

Finally, the first detection of an optical counterpart to one of the
lobes of the inner double in a restarting (double-double) FRII source
implies that the jet termination in these newly formed lobes is also
capable of generating the type of strong shock expected to be
responsible for high-energy particle acceleration, which is in
contrast to the expectations of some models for the inner lobes
\citep[e.g.][]{Brocksopp+11}. This opens up potentially a new window
for understanding the nature of these short-lived inner structures in
restarting systems.

\subsection{Outlook for the future}

Future work, combining radio and \Euclid\ data and spectroscopy
provided by WEAVE-LOFAR, will give us
statistically robust samples to address the question of what drives
the high-energy particle acceleration in hotspots. We can make
progress in two distinct ways.

Firstly, in the EDF-N itself, we expect the optical and IR
sensitivities to increase by a factor of 6 over the lifetime of \Euclid, while the radio data sensitivity
will also increase very substantially as much more of the data that we
have in hand is processed. This will allow us to improve the
constraints on the optical and IR flux densities of the hotspots of sources
already described in this paper, potentially including detections of
as yet undetected objects. Full-resolution LOFAR data over the complete
EDF-N area would be very helpful for this effort, but would require
additional LOFAR observations. Visual or automated searches of
high-resolution LOFAR data for FRIIs over the whole EDF-N are expected
to increase the number of targets significantly; in this case,
selection of candidate FRIIs would be best done by an automated method
\citep[e.g.][]{Clews+25} 
rather than by visual inspection.

Secondly, the Euclid Wide Survey will eventually cover around
14\,000\,deg$^2$ of the extragalactic sky at the depth of the current
Q1 release, the northern half of which will largely be covered by
already existing LoTSS DR3 radio data at 6\arcsecond\ resolution,
while the southern half will be covered by the EMU survey
\citep{Hopkins+25} which will be delivered on a timescale comparable
to the \Euclid\ surveys themselves. A large fraction of the wide-area
\Euclid\ sky will also be visible to VLASS, albeit with lower
sensitivity. It is already the intention of the LOFAR surveys team to
generate a wide-area radio-optical cross-match based on the
\Euclid\ optical catalogues, and other similar efforts will be carried
out by other teams. Since this data set will contain millions of radio
AGN, including at least 200\,000 FRII radio galaxies and
quasars (figures estimated from the LoTSS DR2 catalogue of
\citealt{Hardcastle+25}), a byproduct will be exquisitely good
constraints on particle acceleration to high energies in hotspots,
particularly when combined with ground-based ancillary data to improve
constraints on the SED.

The detection of a small fraction of visually
selected FRIIs in the Euclid Q1 data indicates that large-scale
studies are viable, opening a new window on the physics of such
sources on a statistical basis; again, automated selection of
candidate FRIIs will be essential to allow the hotspots to be selected
efficiently. Optimal constraints will be achieved by matching
\Euclid\ images to the high-resolution data to be generated from the
LOFAR2.0 survey ILoTSS, which will have around 5200\,deg$^2$ of
overlap with \Euclid\ at a target resolution of \mbox{\ang{;;0.3}}. As
we have shown above, cross-matching with radio data at a resolution
well matched to that of the optical and IR gives us the most robust
selection possible of the optical counterparts of radio features.

\section*{Data availability}

Table A.1 is only available in electronic form at the CDS via anonymous ftp to cdsarc.u-strasbg.fr (130.79.128.5) or via \url{http://cdsweb.u-strasbg.fr/cgi-bin/qcat?J/A+A/}.

\begin{acknowledgements}
MJH thanks the UK STFC for support [ST/V000624/1, ST/Y001249/1]. LKM is grateful for support from a UKRI FLF [MR/Y020405/1] and STFC for LOFAR-UK [ST/V002406/1]. Co-funded by the European Union (MSCA Doctoral Network EDUCADO, GA
101119830 and Widening Participation, ExGal-Twin, GA 101158446). JHK
acknowledges grant PID2022-136505NB-I00 funded by
MCIN/AEI/10.13039/501100011033 and EU, ERDF. We thank an
  anonymous referee for constructive comments on the paper.

  \AckQone
  \AckDatalabs
  \AckEC

LOFAR is the Low Frequency Array, designed and constructed by ASTRON.
It has observing, data processing, and data storage facilities in
several countries, which are owned by various parties (each with their
own funding sources), and which are collectively operated by the LOFAR
ERIC under a joint scientific policy. The LOFAR resources have benefited
from the following recent major funding sources: CNRS-INSU,
Observatoire de Paris and Université d'Orléans, France; BMBF,
MIWF-NRW, MPG, Germany; Science Foundation Ireland (SFI), Department
of Business, Enterprise and Innovation (DBEI), Ireland; NWO, The
Netherlands; The Science and Technology Facilities Council, UK;
Ministry of Science and Higher Education, Poland; and The Istituto
Nazionale di Astrofisica (INAF), Italy.

This research made use of the University of Hertfordshire
high-performance computing facility and the LOFAR-UK computing
facility located at the University of Hertfordshire (\url{https://uhhpc.herts.ac.uk}) and supported by
STFC [ST/P000096/1].

This research made use of \software{Astropy}, a community-developed core
Python package for astronomy \citep{AstropyCollaboration13} hosted at
\url{http://www.astropy.org/}, of \software{Matplotlib} \citep{Hunter07},
of \software{APLpy}, an open-source astronomical plotting package for
Python hosted at \url{http://aplpy.github.com/}, and of \software{topcat}
\citep{Taylor05}.

The National Radio Astronomy Observatory is a facility of the National
Science Foundation operated under cooperative agreement by Associated
Universities, Inc.

This research used data obtained with the Dark Energy Spectroscopic Instrument (DESI). DESI construction and operations is managed by the Lawrence Berkeley National Laboratory. This material is based upon work supported by the U.S. Department of Energy, Office of Science, Office of High-Energy Physics, under Contract No. DE–AC02–05CH11231, and by the National Energy Research Scientific Computing Center, a DOE Office of Science User Facility under the same contract. Additional support for DESI was provided by the U.S. National Science Foundation (NSF), Division of Astronomical Sciences under Contract No. AST-0950945 to the NSF’s National Optical-Infrared Astronomy Research Laboratory; the Science and Technology Facilities Council of the United Kingdom; the Gordon and Betty Moore Foundation; the Heising-Simons Foundation; the French Alternative Energies and Atomic Energy Commission (CEA); the National Council of Humanities, Science and Technology of Mexico (CONAHCYT); the Ministry of Science and Innovation of Spain (MICINN); and by the DESI Member Institutions: \url{https://www.desi.lbl.gov/collaborating-institutions}. The DESI collaboration is honoured to be permitted to conduct scientific research on I’oligam Du’ag (Kitt Peak), a mountain with particular significance to the Tohono O’odham Nation. Any opinions, findings, and conclusions or recommendations expressed in this material are those of the author(s) and do not necessarily reflect the views of the U.S. National Science Foundation, the U.S. Department of Energy, or any of the listed funding agencies.

\end{acknowledgements}

\bibliography{unified} 

\label{LastPage} 
\clearpage
\end{document}